\pdfoutput=1

\documentclass[prd, twocolumn, superscriptaddress,floatfix, nofootinbib]{revtex4-2}

\usepackage{url}
\usepackage{xspace}
\usepackage{dsfont}
\usepackage{amssymb}
\usepackage{amsmath}
\usepackage{graphicx}
\graphicspath{ {./} }
\usepackage[small]{subfigure}
\usepackage[colorlinks=true,citecolor=blue]{hyperref}
\usepackage{multirow}
\usepackage{hhline}
\usepackage{color}
\definecolor{darkred}{rgb}{1.0,0.1,0.1}
\definecolor{darkgreen}{rgb}{0.1,0.7,0.1}
\definecolor{darkblue}{rgb}{0.1,0.1,1.0}

\usepackage{amsmath}
\DeclareMathOperator*{\argmax}{argmax}
\DeclareMathOperator*{\argmin}{argmin}

\def\alphas{\texttt{TimeShower:alphaSvalue}}
\def\lund{\texttt{StringZ:aLund}}
\def\strange{\texttt{StringFlav:probStoUD}}

\begin{document}

\title{Parameter Estimation using Neural Networks in the Presence of Detector Effects}

\author{Anders Andreassen}
\email{ajandreassen@google.com}
\affiliation{Google, Mountain View, CA 94043, USA}

\author{Shih-Chieh Hsu}
\email{schsu@uw.edu}
\affiliation{Department of Physics, University of Washington, Seattle, Washington 98195, USA}

\author{Benjamin Nachman}
\email{bpnachman@lbl.gov}
\affiliation{Physics Division, Lawrence Berkeley National Laboratory, Berkeley, CA 94720, USA}
\affiliation{Berkeley Institute for Data Science, University of California, Berkeley, CA 94720, USA}

\author{Natchanon Suaysom}
\email{nsuaysom@uw.edu}
\affiliation{Department of Physics, University of Washington, Seattle, Washington 98195, USA}

\author{Adi Suresh}
\email{adisurtya@berkeley.edu}
\affiliation{Physics Division, Lawrence Berkeley National Laboratory, Berkeley, CA 94720, USA}
\affiliation{Department of Physics, University of California, Berkeley, CA 94720, USA}

\begin{abstract}
Histogram-based template fits are the main technique used for estimating parameters of high energy physics Monte Carlo generators.  Parametrized neural network reweighting can be used to extend this fitting procedure to many dimensions and does not require binning.  If the fit is to be performed using reconstructed data, then expensive detector simulations must be used for training the neural networks.  We introduce a new two-level fitting approach that only requires one dataset with detector simulation and then a set of additional generation-level datasets without detector effects included.  This \textbf{S}imulation-level fit based on \textbf{R}eweighting \textbf{G}enerator-level events with \textbf{N}eural networks (\textsc{Srgn}) is demonstrated using simulated datasets for a variety of examples including a simple Gaussian random variable, parton shower tuning, and the top quark mass extraction.
\end{abstract}

\maketitle

\section{Introduction}

Synthetic data produced from Monte Carlo (MC) generators are a key tool for statistical analysis in high energy particle physics.  These MC generators have a number of parameters that can be measured by producing multiple synthetic datasets and finding the one that agrees best with data.  This procedure can be computationally expensive, especially when detector simulations are involved.  In some cases, one can avoid detector simulations by using unfolded data for parameter estimation.  Until recently~\cite{Andreassen:2019cjw,Bellagente:2019uyp}, unfolding methods were only available for low dimensional or binned data.  Even with the advent of new methods, one can achieve a higher precision with folding instead of unfolding.  For example, template-based fitting is the standard approach for extracting the top quark mass~\cite{ATLAS:2014wva}, one of the most precisely measured quantities at hadron colliders\footnote{Even with $\sim 1$ GeV theoretical ambiguity~\cite{Hoang:2020iah,Corcella:2019tgt}, the uncertainty is still at the $0.5\%$ level.}.

Machine learning may provide a solution to the simulation challenge.  One possibility is to replace or augment synthetic data from slow physics-based generators with synthetic data generated from neural generative models~\cite{Paganini:2017hrr,Paganini:2017dwg,Vallecorsa:2019ked,Chekalina:2018hxi,ATL-SOFT-PUB-2018-001,Carminati:2018khv,Vallecorsa:2018zco,Musella:2018rdi,Erdmann:2018kuh,Erdmann:2018jxd,Oliveira:DLPS2017,deOliveira:2017rwa,Hooberman:DLPS2017,Belayneh:2019vyx,Buhmann:2020pmy,deOliveira:2017pjk,Butter:2019eyo,Martinez:2019jlu,Bellagente:2019uyp,Vallecorsa:2019ked,SHiP:2019gcl,Carrazza:2019cnt,Butter:2019cae,Lin:2019htn,DiSipio:2019imz,Hashemi:2019fkn,Zhou:2018ill,Datta:2018mwd,Deja:2019vcv,Derkach:2019qfk,Erbin:2018csv,Urban:2018tqv,Farrell:2019fsm}.  This requires neural networks to learn $p(\text{data}|\text{parameters})$ accurately, which is a difficult task.  An alternative solution is to instead learn the ratio $p(\text{data}|\text{parameters})/p(\text{data}|\text{reference})$, where the reference may be from a particular synthetic dataset generated with a fixed set of parameters.  It is well-known~\cite{hastie01statisticallearning,sugiyama_suzuki_kanamori_2012} (also in high energy physics~\cite{Andreassen:2019nnm,Badiali:2020wal,Stoye:2018ovl,Hollingsworth:2020kjg,Brehmer:2018kdj,Brehmer:2018eca,Brehmer:2019xox,Brehmer:2018hga,Cranmer:2015bka,Badiali:2020wal,Andreassen:2020nkr,Andreassen:2019cjw,Fischer-ACAT2019}) that a suitably structured and trained neural network-based classifier learns to approximate this likelihood ratio, so one can turn the difficult problem of probability density estimation into the relatively easier task of classification.  Applying this idea to full phase space reweighting and parameter estimation was recently proposed with the \textit{Deep neural networks using Classification for Tuning and Reweighting} (\textsc{Dctr}) protocol~\cite{Andreassen:2019nnm}.  When used to perform an unbinned fit, the original \textsc{Dctr} algorithm first learns a parametrized reweighting function and then continuously (and differentially) modifies the MC generator parameters until the classifier loss used to define the reweighting function is minimized.

The \textsc{Dctr} fitting protocol is effective because it factorizes the reweighting and fitting steps.  Furthermore, the fit can be performed with gradient-based methods due to the differentiability of neural networks.  However, a key challenge with this approach is that one must train the reweighting function using data of the same type as the data that are used in the fit.  In other words, if the fit is performed with data at detector-level, the reweighting function must be trained with a large number of synthetic data examples that include detector effects.  As detector simulations can be computationally expensive, this can be a significant challenge.

We propose a new approach whereby only one synthetic dataset with detector effects ("simulation") is required and all of the reweighting is performed at particle level ("generation") (following the nomenclature from Ref.~\cite{Andreassen:2019cjw}).  This new \textbf{S}imulation-level fit based on \textbf{R}eweighting \textbf{G}enerator-level events with \textbf{N}eural networks (\textsc{Srgn}) approach still factorizes the problem into a reweighting step and a fitting step, except that now each step includes training classifiers: one at generator level and one at simulation level, respectively.  This approach is the same as  \textsc{Dctr} in the reweighting step but differs in the fitting step.  In the form proposed in this paper, the fitting step is not differentiable, but it is amenable to nongradient-based optimization procedures.  Given the computational efficiency of particle-level generation compared with detector-level simulation, this approach will enable new fitting strategies for analyses like the top quark mass measurement, related tasks at the Large Hadron Collider (LHC), and beyond.

Computationally, this new approach requires an up-front cost to train a continuous reweighting function and then a per-evaluation cost to train a classifier to determine a goodness-of-fit.  The fitting step could use a computationally cheaper and more standard goodness-of-fit metric at the cost of precision.  This is more computationally effective than the traditional approach, which requires many simulated samples at detector level.

This paper is organized as follows.  Section~\ref{sec:theory} reviews neural network reweighting and introduces the new two-level approach for incorporating detector effects.  A variety of numerical results are presented in Sec.~\ref{sec:results}.  In particular, (1) a simple Gaussian example is used to first demonstrate the salient features of the new approach, then (2) parton shower tuning provides a high-dimensional example without detector effects, and finally (3) the top quark mass measurement is deployed for a multidimensional use case including detector effects.  The paper ends with conclusions and outlook in Sec.~\ref{sec:conclusions}.

\section{Neural Network Reweighting and Detector Effects}
\label{sec:theory}

Suppose that features $X\in\mathbb{R}^N$ follow a probability density $p(x|\theta)$, where $\theta$ are parameters of the model.  A reweighting function $w_{\theta_0}(x,\theta)$ is designed so that a sample drawn from $p(x|\theta_0)$ weighted by $w$ is statistically identical to a sample drawn from $p(x|\theta)$.  The ideal reweighting function is $w_{\theta_0}(x,\theta)=p(x|\theta)/p(x|\theta_0)$.  One strategy for constructing $w$ is to model the probability density $p(x|\theta)$ and then take the ratio\footnote{This was used in a recently proposed parametric unfolding method in Ref.~\cite{Gagunashvili:2020srr}.}.  Density estimation is a significant challenge, especially in the case of collision events where $X$ is a variable and high-dimensional object and $p(x)$ has significant symmetry.  One solution is to turn the challenge of density estimation into the relatively easier task of classification.  Suppose that $f$ is a neural network trained to distinguish between a sample of events $\boldsymbol{\theta}$ drawn from $p(x|\theta)$ and a sample of events $\boldsymbol{\theta_0}$ drawn from $p(x|\theta_0)$.  If $f$ is trained using the binary cross entropy loss function:
\begin{align}
    \text{Loss}(f(x))=-\sum_{x_i\in\boldsymbol{\theta}}\log(f(x_i))-\sum_{x_i\in\boldsymbol{\theta_0}}\log(1-f(x_i))\,,
\end{align}
then with a flexible enough architecture, an effective training protocol, and sufficient training data, 
\begin{align}
    \frac{f(x)}{1-f(x)}\propto\frac{p(x|\theta)}{p(x|\theta_0)}\,.
\end{align}
Therefore, one can construct $w$ using $f$.  Furthermore, if the training of $f$ includes a variety of values of $\theta$, then it will naturally learn to interpolate and become $f(x,\theta)$; consequently, $w$ becomes a parametrized reweighting function.

The original \textsc{Dctr} fitting protocol is expressed symbolically as

\begin{align}
\nonumber
    \theta_{\textsc{Dctr}}^*&=\argmax_{\theta'}\sum_{x_i\in\boldsymbol{\theta_?}}\log(f(x_i,\theta'))\\\label{eq:dctrfit}
    &\hspace{10mm}+\sum_{x_i\in\boldsymbol{\theta_0}}\log(1-f(x_i,\theta'))\,,
\end{align}
where $\theta_?$ is not known.  If $f$ is the optimal classifier, then $\theta_{\textsc{Dctr}}^*=\theta_{?}$.

Detector effects distort the feature space.  Let $X_S|X_G\in \mathbb{R}^M$ represent simulation-level features given generator-level features $X_G$.  In synthetic samples, 
we have the corresponding pairs of $X_G$ and $X_S$ for every collision event.  However, $X_G$ is not known for real data.  Therefore, it would be ideal to do the fit using $X_S$, but perform the reweighting using $X_G$, as reweighting only requires events from generation.  

\begin{figure}[h!]
    \centering
    \includegraphics[width=0.45\textwidth]{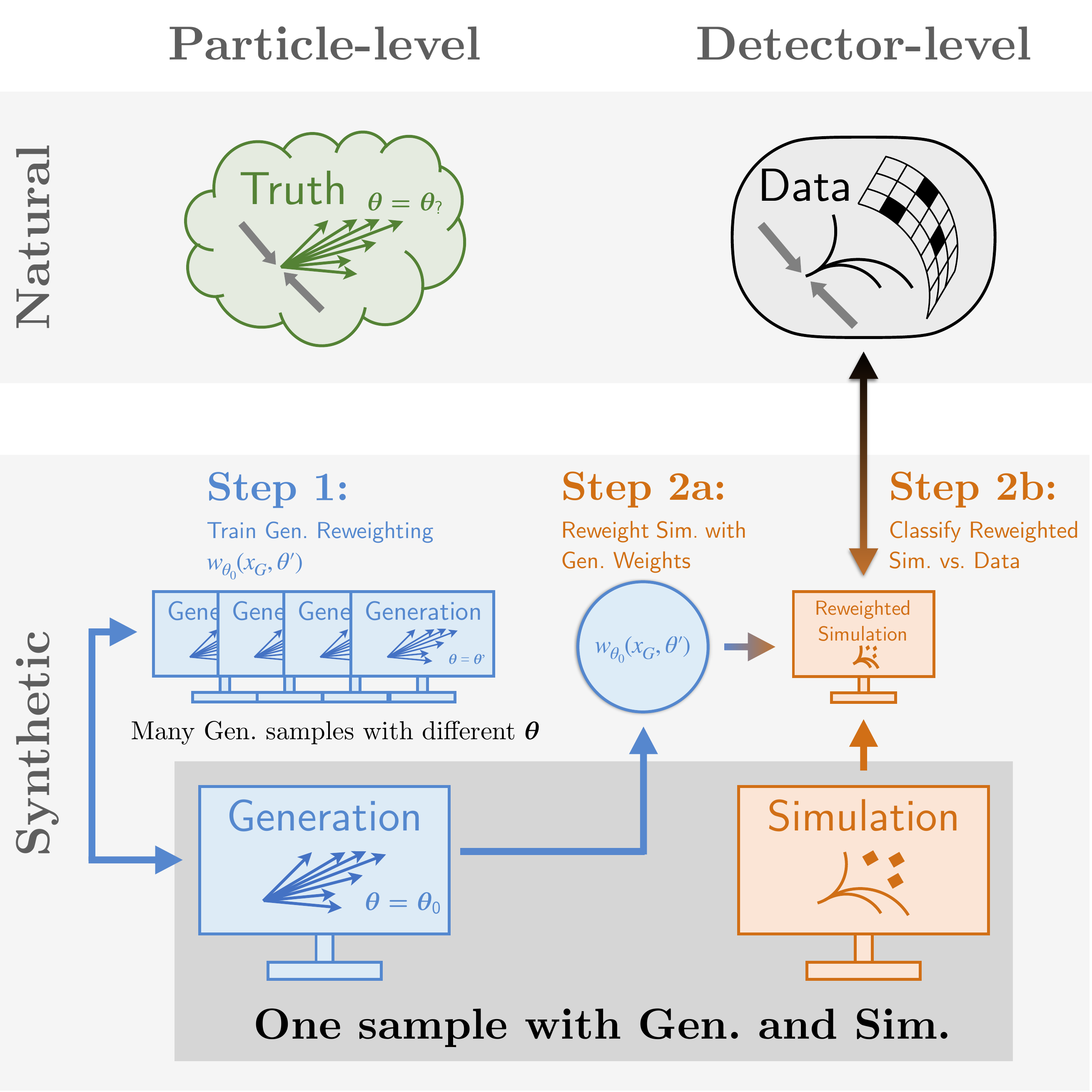}
    \caption{An illustration of \textsc{Srgn}, applied to a set of synthetic and natural data.  There is one synthetic dataset where the particle-level data (generation) is passed through a detector emulation (simulation).  \textsc{Srgn} is a two-step process. First, a parametrized reweighting function is learnt using the generation dataset and a set of additional synthetic generator-level datasets. Second, the synthetic simulation is reweighted and compared with real data, iterated to converge on the parameters $\theta_?$. Illustration is inspired by Ref.~\cite{Andreassen:2019cjw}.}
    \label{fig:my_label}
\end{figure}

The \textsc{Srgn} protocol is a two-step procedure as illustrated in Fig.~\ref{fig:my_label}.  First, a reweighting function is trained.  Then, a classifier is trained to distinguish the target data from reweighted simulation.  As this classifier is trained, the parameters $\theta$ are also modified.  When the classifier is unable to distinguish the target data from the reweighted simulation, then the current parameters are the fitted parameters.

Symbolically, suppose that $w_{\theta_0}(x_{G},\theta)$ is a reweighting function learned at generator level, where $\theta_0$ represents the nominal parameters for the synthetic sample.  Furthermore, suppose that $g$ is a neural network defined as follows:

\begin{align}\nonumber
    g_{\theta'}(x_S)&=\argmax_{g}\sum_{x_{S,i}\in\boldsymbol{\theta_?}}\log(g_{\theta'}(x_{S,i}))+\\\label{eq:gdef}
    &\hspace{10mm}\sum_{(x_{G,i},x_{S,i})\in\boldsymbol{\theta_0}}w_{\theta_0}(x_{G,i},\theta')\log(1-g_{\theta'}(x_{S,i}))\,.
\end{align}
Then,
\begin{align}
\label{eq:srgnfit}
    \theta_{\textsc{Srgn}}^*\equiv\argmin_{\theta'}\left[\Pr(g_{\theta_?}(x_S)>g_{\theta'}(x_S))\right]\,,
\end{align}
where the quantity in $[\cdot]$ is typically called the area under the receiver operating characteristic curve or AUC. We calculate the AUC between $g$’s predictions on events from the unknown sample and  $g$’s predictions on reweighted events from the synthetic sample;  effectively, if we reweight events from the synthetic sample $\boldsymbol{\theta_0}$ to events from $\boldsymbol{\theta’}$, then we calculate the AUC between $g$’s predictions on  $\boldsymbol{\theta'}$ and $g$’s predictions on $\boldsymbol{\theta_?}$.

In analogy to Eq.~\ref{eq:dctrfit}, one might think to define $\theta_{\textsc{Srgn}}^*$ as the value of $\theta'$ that maximizes the loss in Eq.~\ref{eq:gdef}.  This would make the \textsc{Srgn} procedure differentiable in contrast to Eq.~\ref{eq:srgnfit} (the AUC is not generically differentiable).  However, one can show that (see Appendix~\ref{app:derivation})

\begin{align}
\label{eq:g}
    g_{\theta'}(x_S)\approx\frac{p}{\mathbb{E}[w_{\theta_0}(x_{G},\theta')|x_S](1-p)+p}\,,
\end{align}
where $p=\Pr(\theta=\theta_?|x_S)$.  When $w=1$, Eq.~\ref{eq:g} is the usual result that the classifier is simply the probability of the target class given the features.  Plugging Eq.~\ref{eq:g} into Eq.~\ref{eq:gdef} and optimizing with respect to $\theta'$ does not generally result in $\theta^*=\theta_?$ (see Appendix~\ref{sec:app}). 

The \textsc{Srgn} result defined by Eq.~\ref{eq:srgnfit} achieves $\theta_{\textsc{Srgn}}^*=\theta_{?}$ when the features $x_G$ include the \textit{full phase space}, defined below. 

The probability density of the features $x_S$ weighted by $w_{\theta_0}(x_{G},\theta)$ is given by

\begin{align}
    p_\text{weighted}(x_S|\theta,\theta_0)&\equiv \int p(x_S,x_G|\theta_0)\,w_{\theta_0}(x_{G},\theta)\,dx_G\\\label{eq:approx}
    &\approx \int p(x_S,x_G|\theta_0) \, \frac{p(x_G|\theta)}{p(x_G|\theta_0)}\,dx_G\\\label{eq:fullphasespace}
    &=\int p(x_S|x_G,\theta_0)\,p(x_G|\theta)\,dx_G\,,
\end{align}
where the approximation in Eq.~\ref{eq:approx} depends on the fidelity of the neural network optimization.  Equation~\ref{eq:fullphasespace} is equal to $p(x_S|\theta)$ if $p(x_S|x_G,\theta_0)=p(x_S|x_G,\theta)$.  In this case $\theta_{\textsc{Srgn}}^*=\theta_{?}$.  The equality $p(x_S|x_G,\theta_0)=p(x_S|x_G,\theta)$ holds if $x_G$ contains all of the relevant information about the detector response so that changing $\theta$ has no impact on the resolution.  In this case, the feature space is said to contain the \textit{full phase space} (later denoted $\Omega$).  Note that it is common in experimental analyses to perform generator-level reweighting for estimating theoretical modeling uncertainties.  These reweighting schemes typically use histograms and therefore are constrained to one or two-dimensional feature spaces.  The above calculation suggests\footnote{We have only shown that if $x_G$ is full phase space, then the procedure is unbiased.  However, it could happen that $x_G$ could be less than full phase space, but $p(x_S|x_G,\theta)=p(x_S|x_G,\theta')$ still holds.} that this is likely insufficient for an unbiased estimate of the impact on simulation-level quantities.

The various properties of the \textsc{Srgn} method will be illustrated in the next section with a variety of examples.

\section{Results}
\label{sec:results}

Three sets of examples are used to illustrate various aspects of the \textsc{Srgn} method.  First, simple Gaussian examples are used, where the probability density is known and thus the reweighting function can be computed analytically.  The features of \textsc{Srgn} described in the previous section are explored with these examples.  The parton shower examples from Ref.~\cite{Andreassen:2019nnm} are used as a second example.  These examples show how the new method can be effective with high-dimensional features but do not incorporate detector effects.  A measurement of the top quark mass is used as a third example to demonstrate both multivariate fitting and detector effects.

The \textsc{Srgn} protocol calls for two neural networks: one called $f$ that is used to construct the reweighting function $w$ and another called $g$ that is used to perform the fit.  These neural networks are implemented using \textsc{Keras}~\cite{keras} with the \textsc{Tensorflow} backend~\cite{tensorflow} and optimized with \textsc{Adam}~\cite{adam}.  Networks are trained using the binary cross entropy loss function.  The network architectures vary by example and are described below.

\subsection{Gaussian Example}

\begin{figure}[h!]
	      	\includegraphics[scale=0.5]{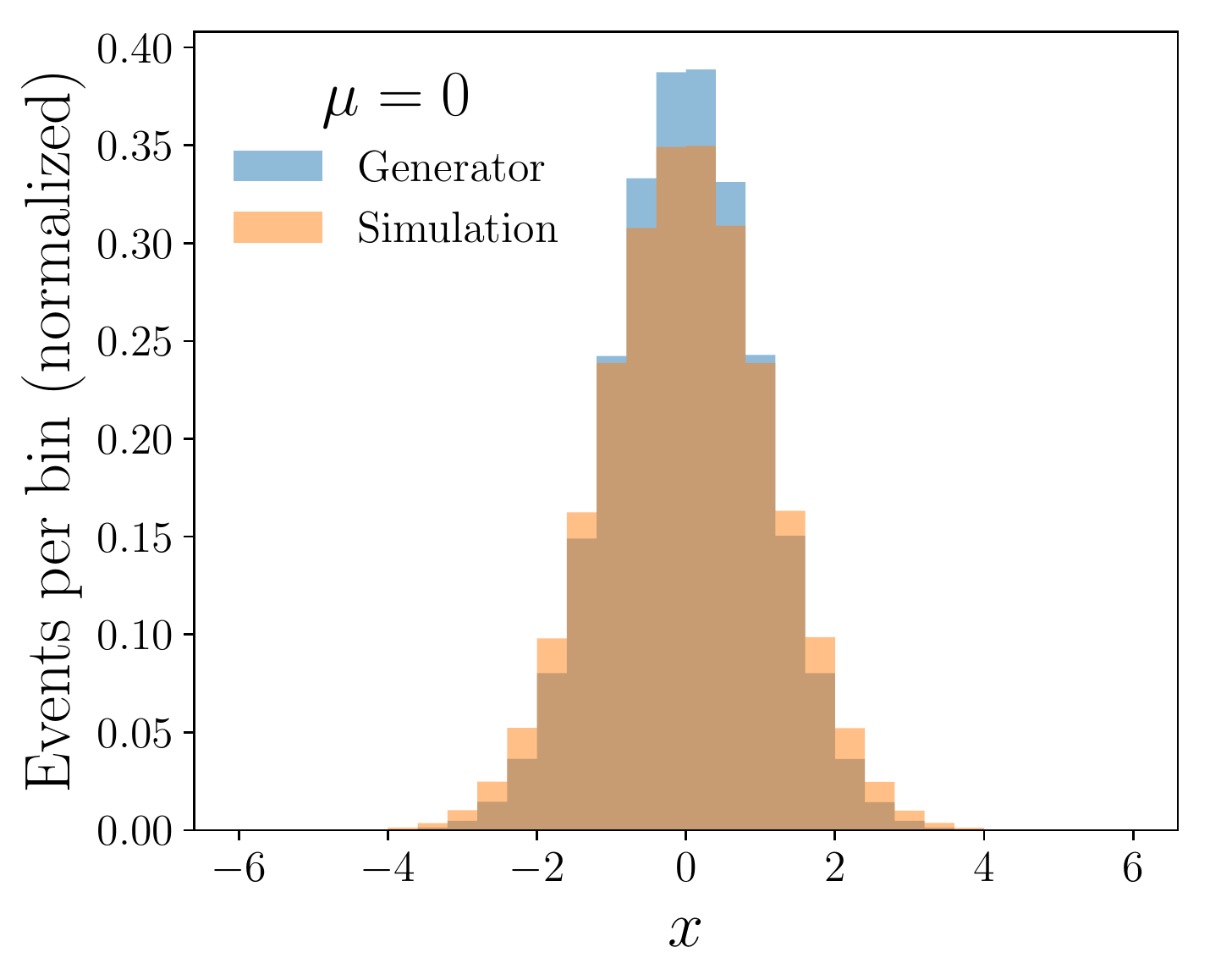}
        	\caption{A histogram of the Gaussian random variable $x$ for $\mu=0$ at generator level and simulation level.}
        	\label{fig:1DGaussianSetup}
        \end{figure}

The generator-level feature space is one-dimensional and follows a Gaussian distribution: $X_G\sim\mathcal{N}(\mu,\sigma^2)$.  Detector effects are modeled as independent Gaussian noise: $X_S=X_G+Z$ where $Z\sim\mathcal{N}(0,\epsilon^2)$.  The detector smearing $\epsilon$ and the generator width $\sigma=1$ are known, but $\mu$ is not known.  In this case, the reweighting function can be computed analytically:

\begin{align}
\label{eq:analyticreweight}
    w_{\mu_0}(x_G, \mu) = \exp\left(\frac{(x_G-\mu_0)^2-(x_G-\mu)^2}{2}\right)\,.
\end{align}
The parametrized reweighting is trained with $\mu$ values sampled uniformly at random in the range $[-2,2]$. One million examples are used for both data and the nominal synthetic dataset, and $\epsilon=0.5$.  These data for $\mu=0$ are presented in Fig.~\ref{fig:1DGaussianSetup}.  

\begin{figure}[h!]
        	\includegraphics[scale=0.5]{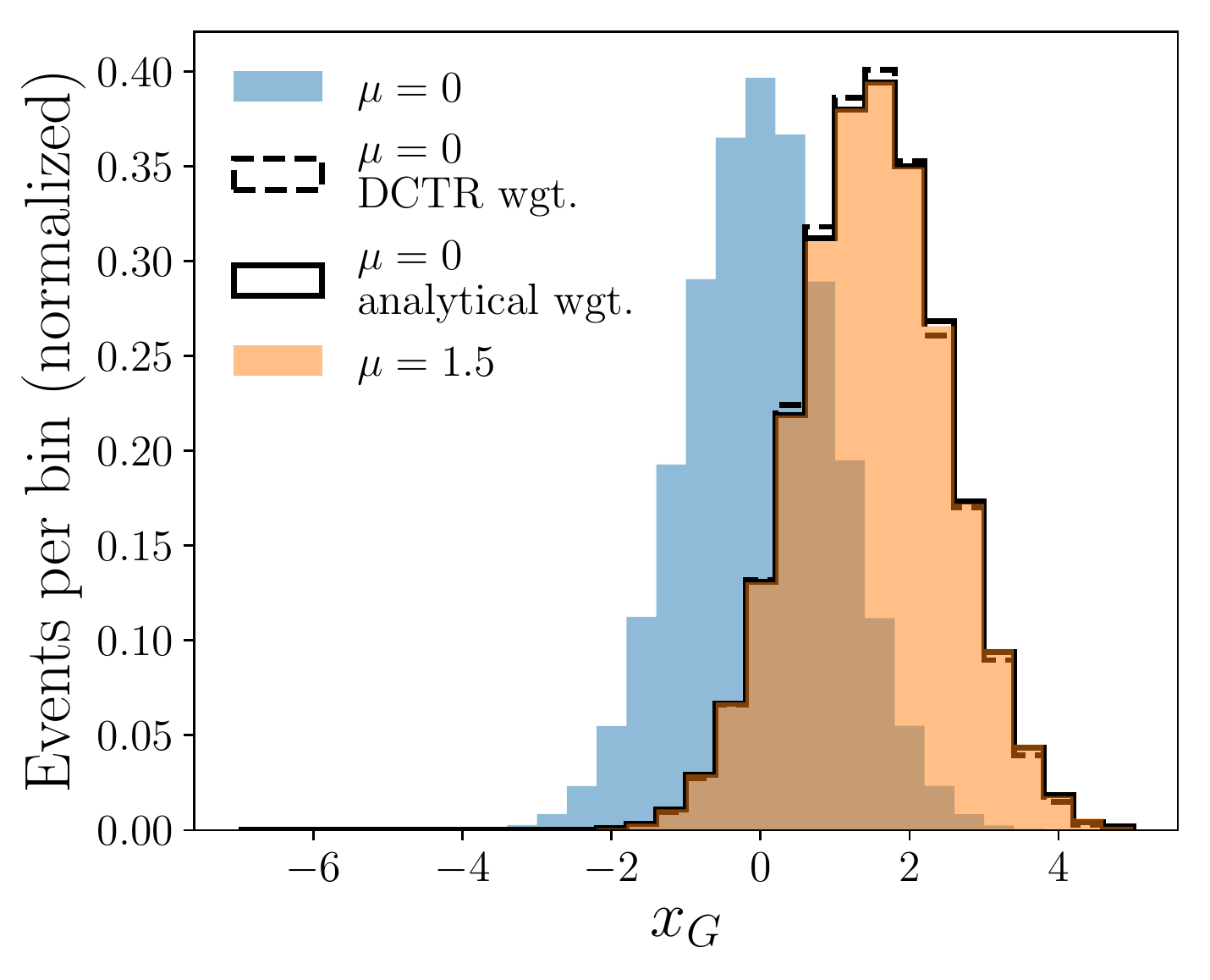}\\
            \includegraphics[scale=0.5]{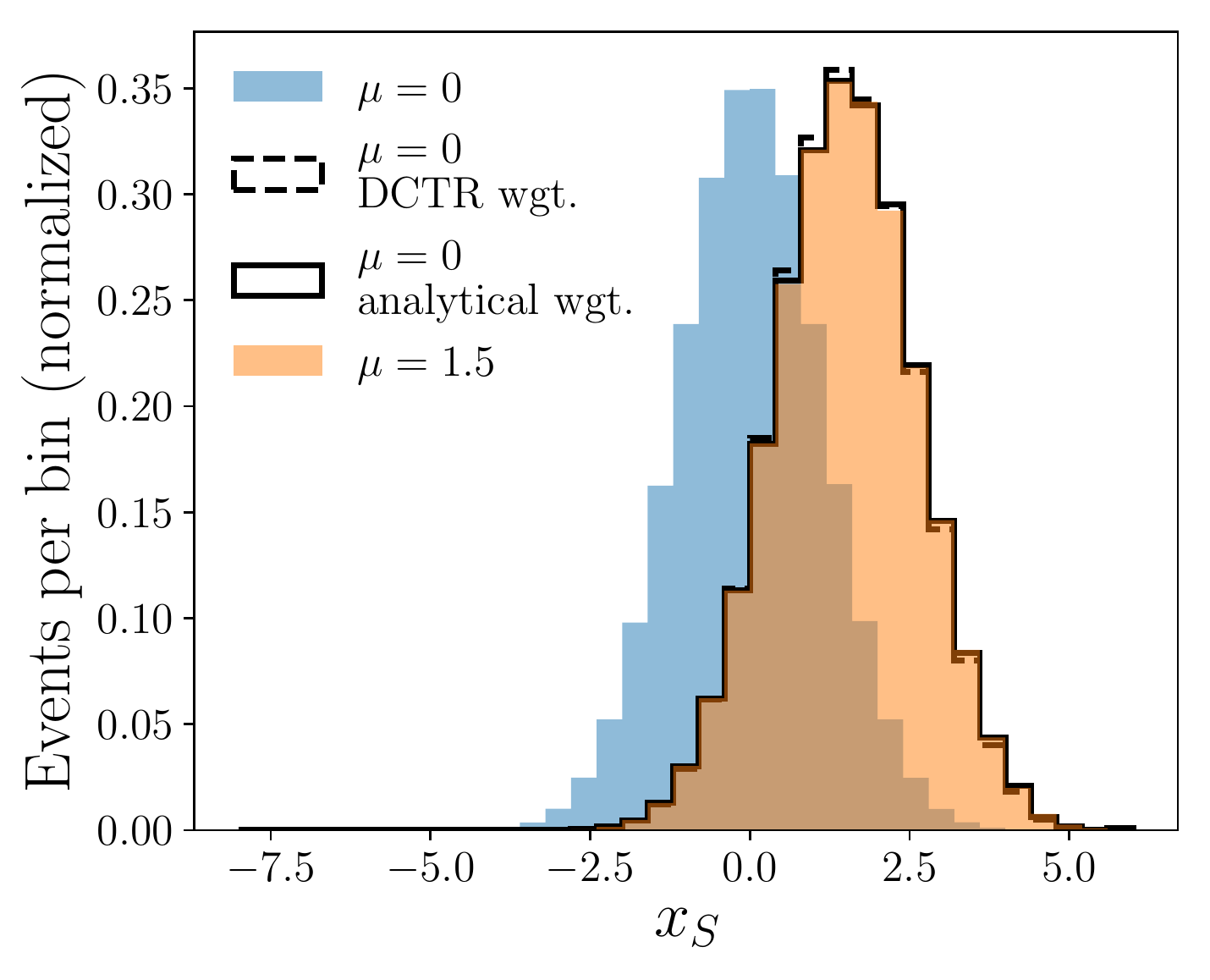}
        	\caption{A demonstration of reweighting for generation (top) and simulation (bottom).}
        	\label{fig:1DGaussianSetup2}
        \end{figure}

A reweighting function is derived using a neural network with three hidden layers using 50 nodes each.  Rectified linear units (ReLU) connect the intermediate layers and the output is Softmax.  The training/validation split is 50/50.  The network is trained for 200 epochs with early stopping using a patience of 10, and the training time is about 5 seconds per epoch on an NVIDIA Tesla V100 ("Volta") GPU. The batch size is $10^5$.  A comparison of the analytical (Eq.~\ref{eq:analyticreweight}) and learned reweighting is shown in Fig.~\ref{fig:1DGaussianSetup2} using weights based on generator level in both cases.  The reweighting effectively morphs the $\mu=0$ distribution to one that has $\mu=1.5$.

\begin{figure}[h!]
	     \includegraphics[scale=0.5]{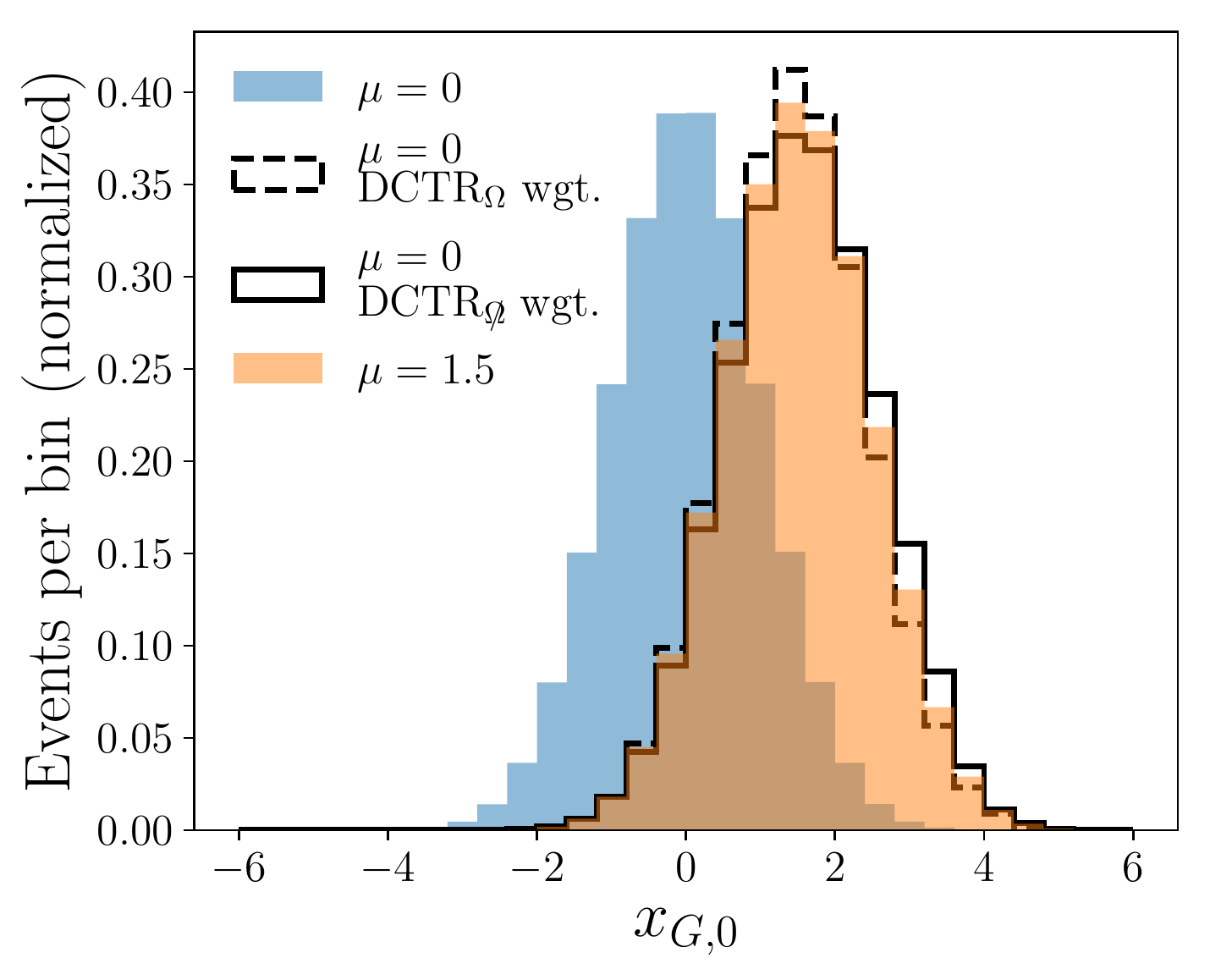}\\
	     \includegraphics[scale=0.5]{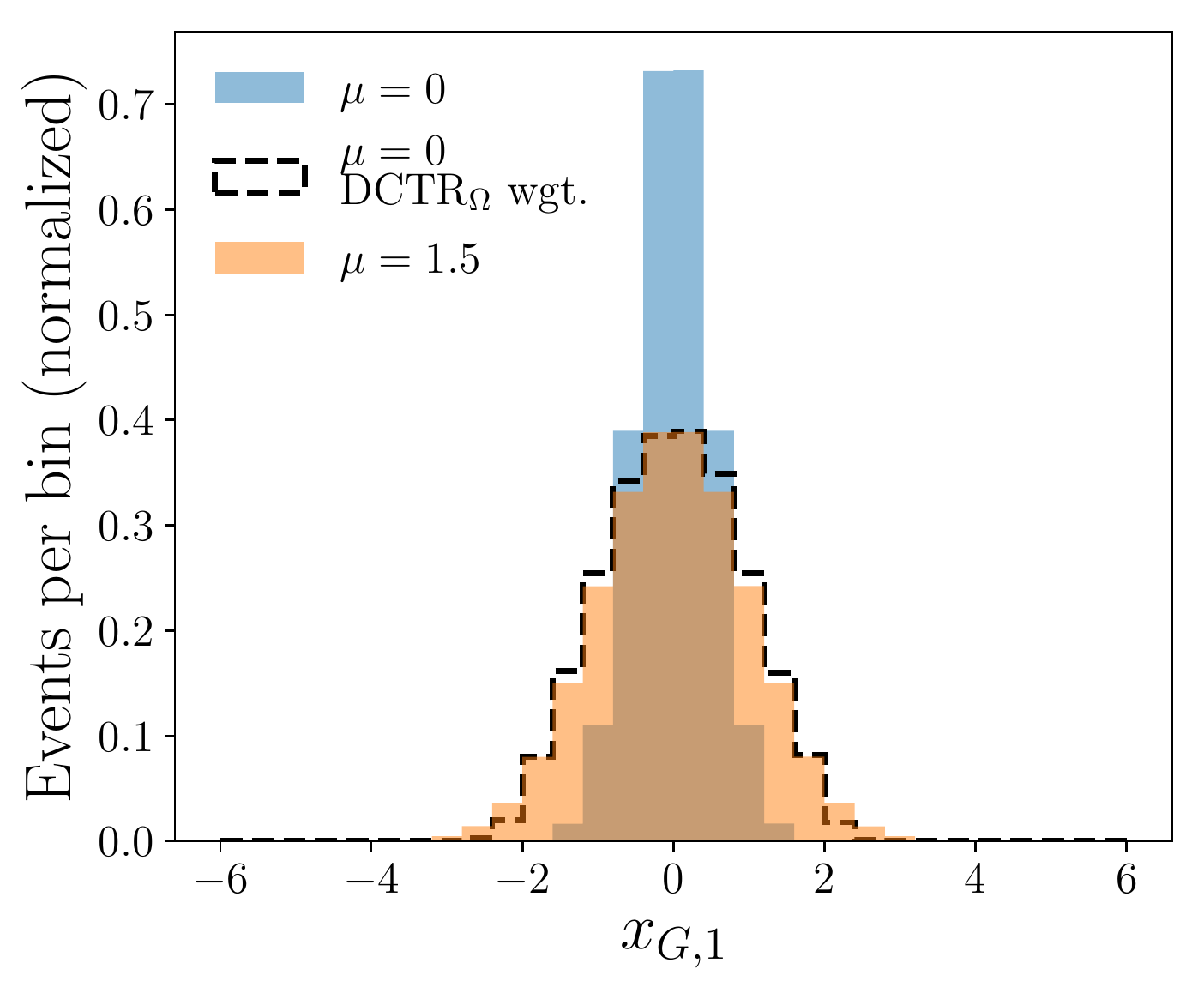}\\
	     \includegraphics[scale=0.5]{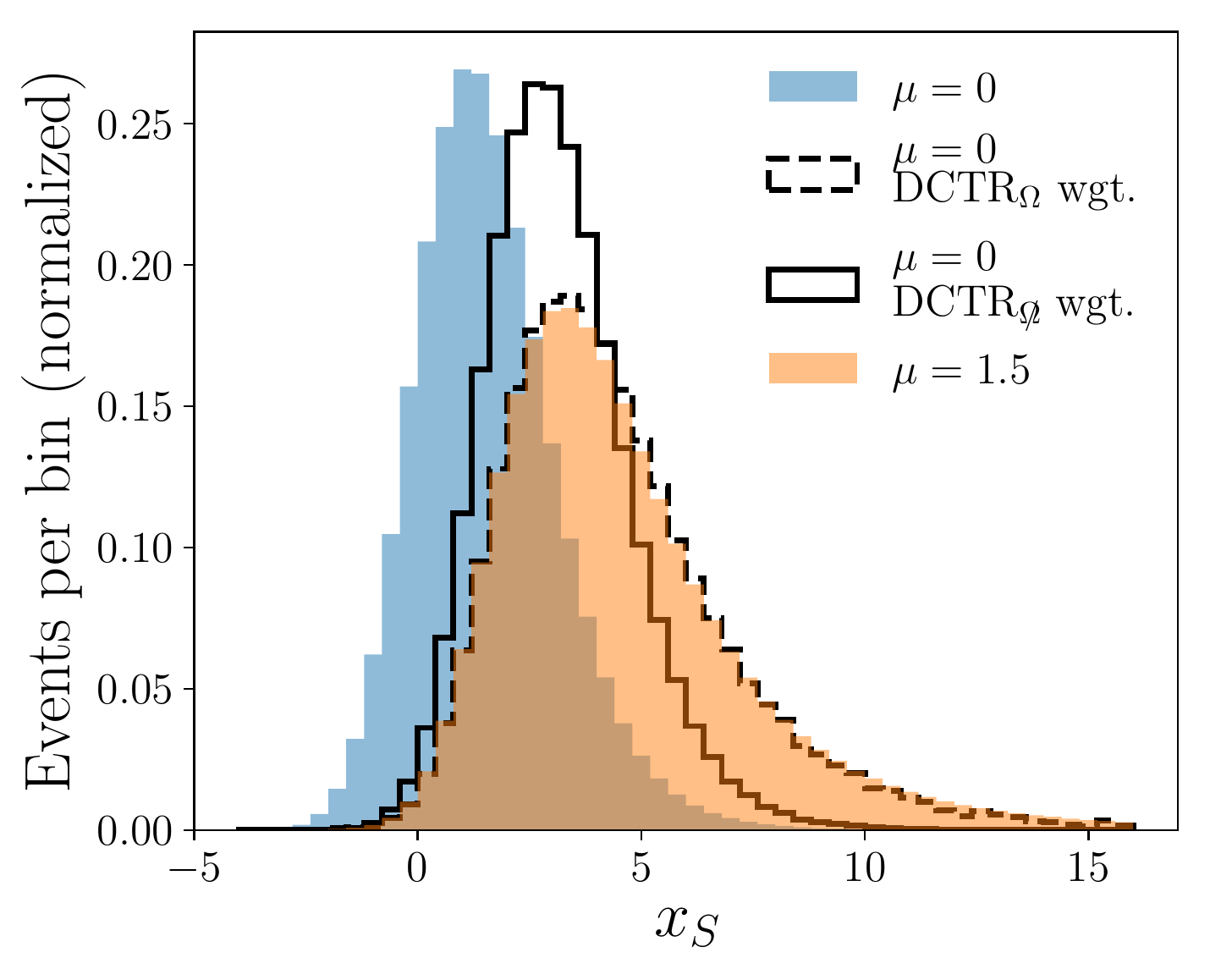}
	     \caption{A demonstration of reweighting derived at generator level for the primary generator-level feature (top), the secondary generator-level feature (middle), and simulation level (bottom).  In the top and bottom plots, a reweighting using only the primary generator feature is also shown (labeled $\text{DCTR}_{\Omega\mkern-10mu/}$ wgt., where "wgt." stands for "reweighted distribution".)
	     }
        \label{fig:reweight2d}
	     \end{figure}

The goal of \textsc{Srgn} is to use simulated features with reweighting based on generator level.  This is explored in Fig.~\ref{fig:reweight2d}.  In order to show that the features need not be the same at generator level and simulation level, $X_G$ is two-dimensional.  Then, we construct the detector response such that the simulation-level observable $X_S$ depends explicitly on the primary generator-level feature, but its detector smearing depends on the secondary generator-level feature.  That is, detector effects are nonuniform, and are dependent on the generator-level parameter(s).  In particular, we choose the primary generator-level feature $X_{G,0}\sim\mathcal{N}(\mu,1)$ and the secondary generator-level feature $X_{G,1}\sim\mathcal{N}(0,\nu^2)$, where  $\nu=(\omega_0 + \omega_1\mu)^2$ for two constants $\omega_0$ and $\omega_1$.  (Specifically, we choose $\omega_0=0.7$ and $\omega_1=0.2$ for this example.)  Then, on a per-event basis, detector effects are emulated by $X_S=X_{G,0}+Z$, where $Z\sim\mathcal{N}(4\lvert x_{G,1}\rvert,(x_{G,1})^4)$, with $4\lvert x_{G,1}\rvert$ representing a net shift bias and $(x_{G,1})^2$ representing a smearing bias.  Importantly, the resolution depends on the secondary generator-level feature.

Figure~\ref{fig:reweight2d} shows the result of a reweighting derived on generator level for ten million events, using the same architecture and training procedure as the previous example.  By construction, both the smearing and the shifting are more intense for the $\mu = 1.5$ simulation-level distribution.  When using both generator-level features (constituting the full phase space $\Omega$), reweighting is successful.  However, if only the primary generator-level feature is used for $w$, then the reweighting fails to reproduce the simulated-level probability density.

\begin{figure}[h!]
	     \includegraphics[scale=0.5]{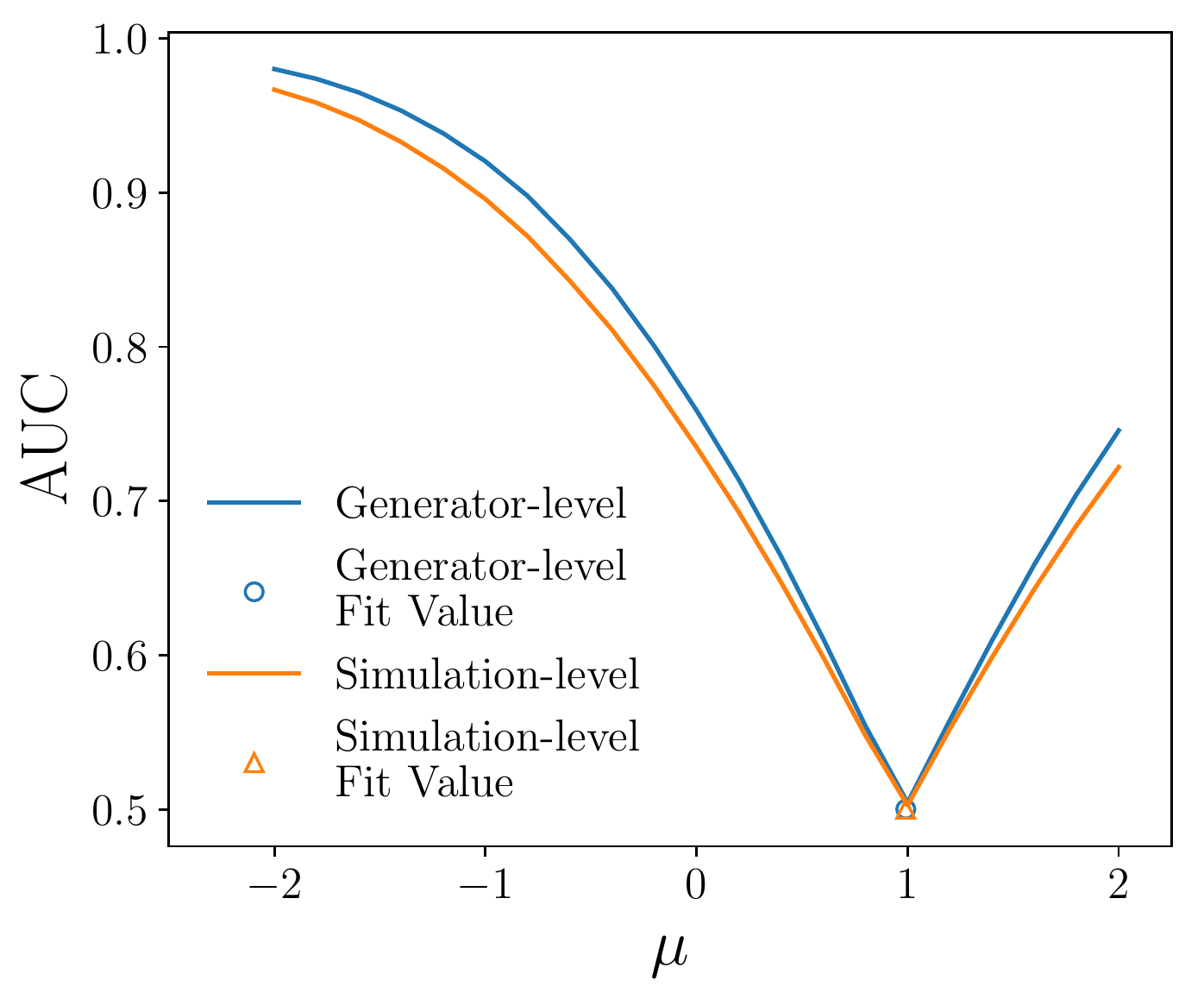}
	     \caption{For an individual run, the AUC versus $\mu$ for generator-level and simulation-level features.  Also shown are the fitted values at both levels using \textsc{SciPy} 1.5.2's \textsc{Powell} nondifferentiable optimizer fitting with 0.0001 as the relative error in both the solution $\mu_{\textsc{Srgn}}$ and the function value $\text{AUC}(\mu_{\textsc{Srgn}})$ acceptable for convergence, within the bounds [-2, 2] (the same range that the reweighting function is parametrized in), and optimizing for a maximum of 100 iterations.}
        \label{fig:gaussian1dauc}
	     \end{figure}

So far, the results have only illustrated the efficacy of reweighting-the rest of the plots in this section demonstrate how the reweighting can be used for fitting.  To begin, the one-dimensional generator-level setup is used for the fit.  The fitting data consist of one million events with $\epsilon=0.5$ for detector effects.  Then, a classifier is trained with different values of $\mu$ to distinguish the unknown dataset from the reweighted synthetic dataset and the AUC from Eq.~\ref{eq:srgnfit} is plotted as a function of $\mu$ for a fit at both generator level and simulation level.  The architecture of this neural network consists of two hidden layers using 128 nodes each.  Rectified linear units (ReLU) connect the intermediate layers and the output is a sigmoid.  The training/validation split is 50/50.  The network is trained for 200 epochs with early stopping using a patience of 5 and the training time is about 5 seconds per epoch on an NVIDIA Tesla V100 GPU.  The batch size is 1000. In both cases, the reweighting is performed at generator level.  Figure~\ref{fig:gaussian1dauc} illustrates several aspects of the proposed fitting method with \textsc{Srgn}.  First, the minimum of the AUC is 0.5 and occurs at $\mu=1$ in both cases, which is the correct value.  Second, the rise in the AUC function away from the minimum is steeper at generator level than simulation level, as expected given the loss of statistical power from detector smearing.  In addition to showing the AUC function, the values of fits using a nondifferentiable optimizer are also presented as markers in Fig.~\ref{fig:gaussian1dauc}.  At both generator level and simulation level, the fit correctly identified $\mu_?=1$.  The numerical results of the one-dimensional Gaussian fit are presented in Table~\ref{tab:gaussian1dfit}, and the reported measurements are the averages and standard deviations over 40 runs using a nondifferentiable optimizer. A small number of runs resulted in defective optimization terminating at the fitting bounds; these were identified and removed by examining values outside a $2\sigma$ window around the mean, as well as all values within 5\% of the range of the bounds.

    \begin{table}[h!]
    	\begin{tabular}{|c|c|c|c|} 
         \hline
         Parameter & Target & Level & Fit value\\ 
         \hline
         \hline
         \multirow{2}{*}{$\mu$} & \multirow{2}{*}{1.000} & Generator & 0.998 $\pm$ 0.013 \\
         & & Simulation &  1.000 $\pm$ 0.017 \\
         \hline
        \end{tabular}
        \caption{Numerical results for the one-dimensional Gaussian fit.  The reported values and errors represent the mean and standard deviation over the 40 runs (with outliers removed), employing the same nondifferentiable optimization scheme described in Fig~\ref{fig:gaussian1dauc}.}
            \label{tab:gaussian1dfit}
    \end{table}

\begin{figure}[h!]
	    \includegraphics[scale=0.5]{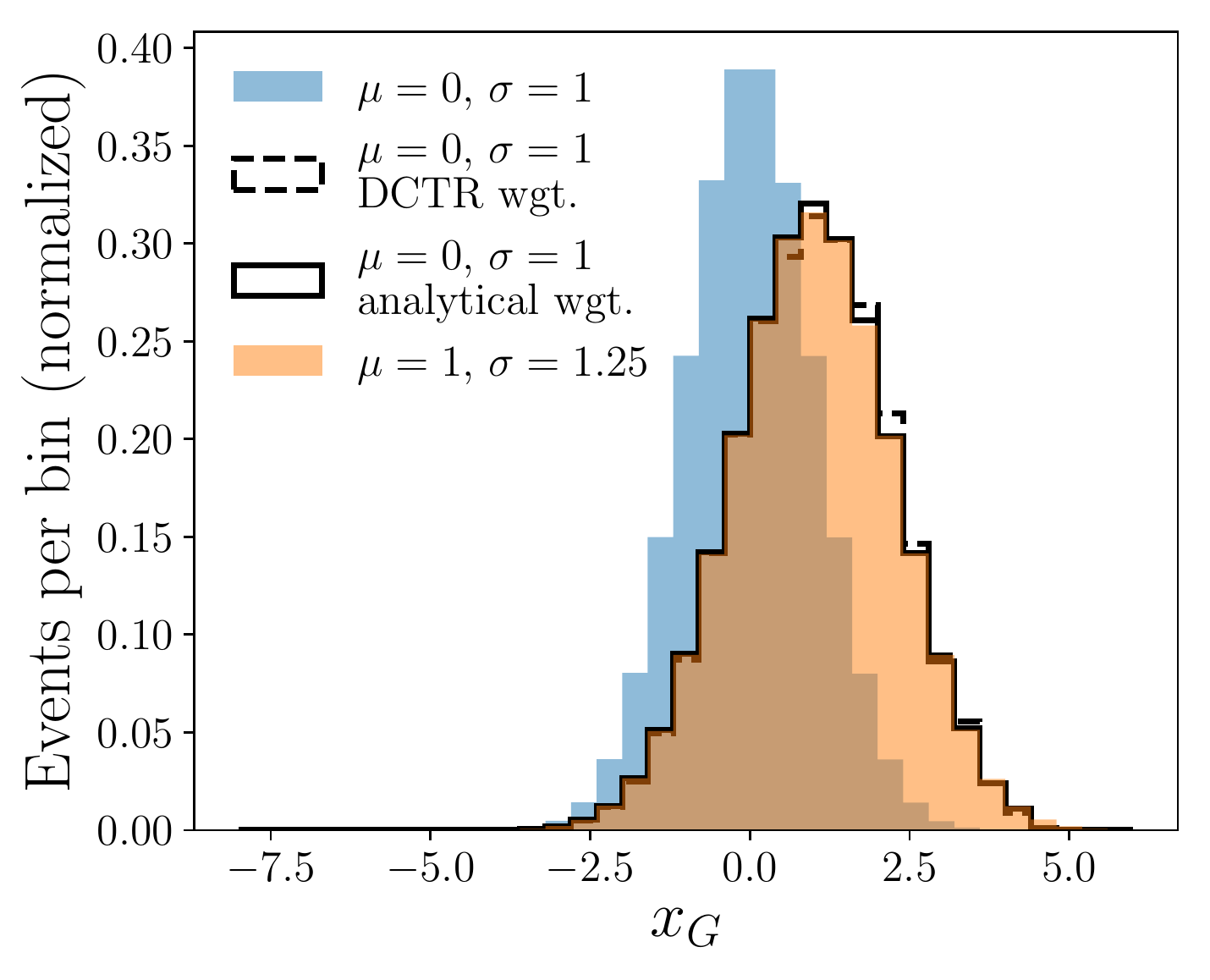}
	   \\
	     \includegraphics[scale=0.5]{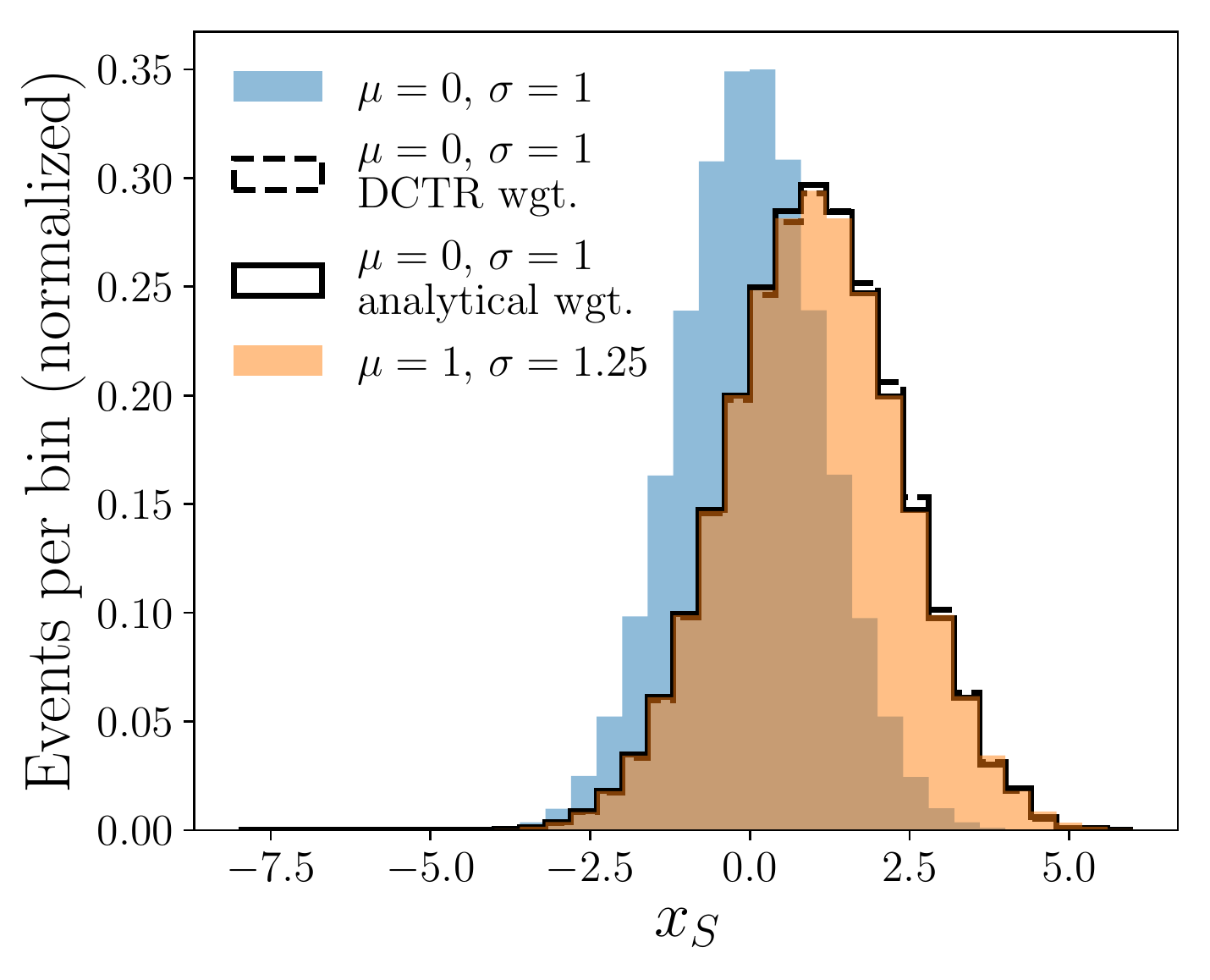}
	     \caption{An illustration of two-dimensional reweighting on generator level (top) and simulation level (bottom).  For comparison, the results with analytical weights from Eq.~\ref{eq:2dgaussiananalytic} and the results with neural network-based weights are both plotted.}
        \label{fig:2dreweight}
	     \end{figure}

As a next illustration, a fit is performed for both $\mu$ and $\sigma$.  A two-dimensional reweighting function is parametrized in the one-dimensional Gaussian mean and standard deviation within the ranges [-2, 2] and [0.25, 4], respectively.  The reweighting function can still be computed analytically and is given by

\begin{align}\nonumber
&w_{(\mu_0,\sigma_0)}(x_G, (\mu, \sigma)) =\\\label{eq:2dgaussiananalytic}
&\hspace{6mm}\frac{\sigma_0}{\sigma}\exp\left(\frac{1}{2}\left(\left(\frac{x_G-\mu_0}{\sigma_0}\right)^2 - \left(\frac{x_G-\mu}{\sigma}\right)^2\right)\right)\,,
\end{align}
where $\mu_0$ and $\sigma_0$ denote the nominal values for the Gaussian distribution.  As before, one million events are used for the fit and detector effects are modeled with $\epsilon=0.5$.  The efficacy of a two-dimensional reweighting function is presented in Fig.~\ref{fig:2dreweight} for a case with $\mu_0=0$ and $\sigma_0=1$.  The neural network weights are just as effective as the analytic weights to morph the default distribution into a distribution with $\mu=1$ and $\sigma=1.25$.

\begin{figure}[h!]
	  \includegraphics[scale=0.5]{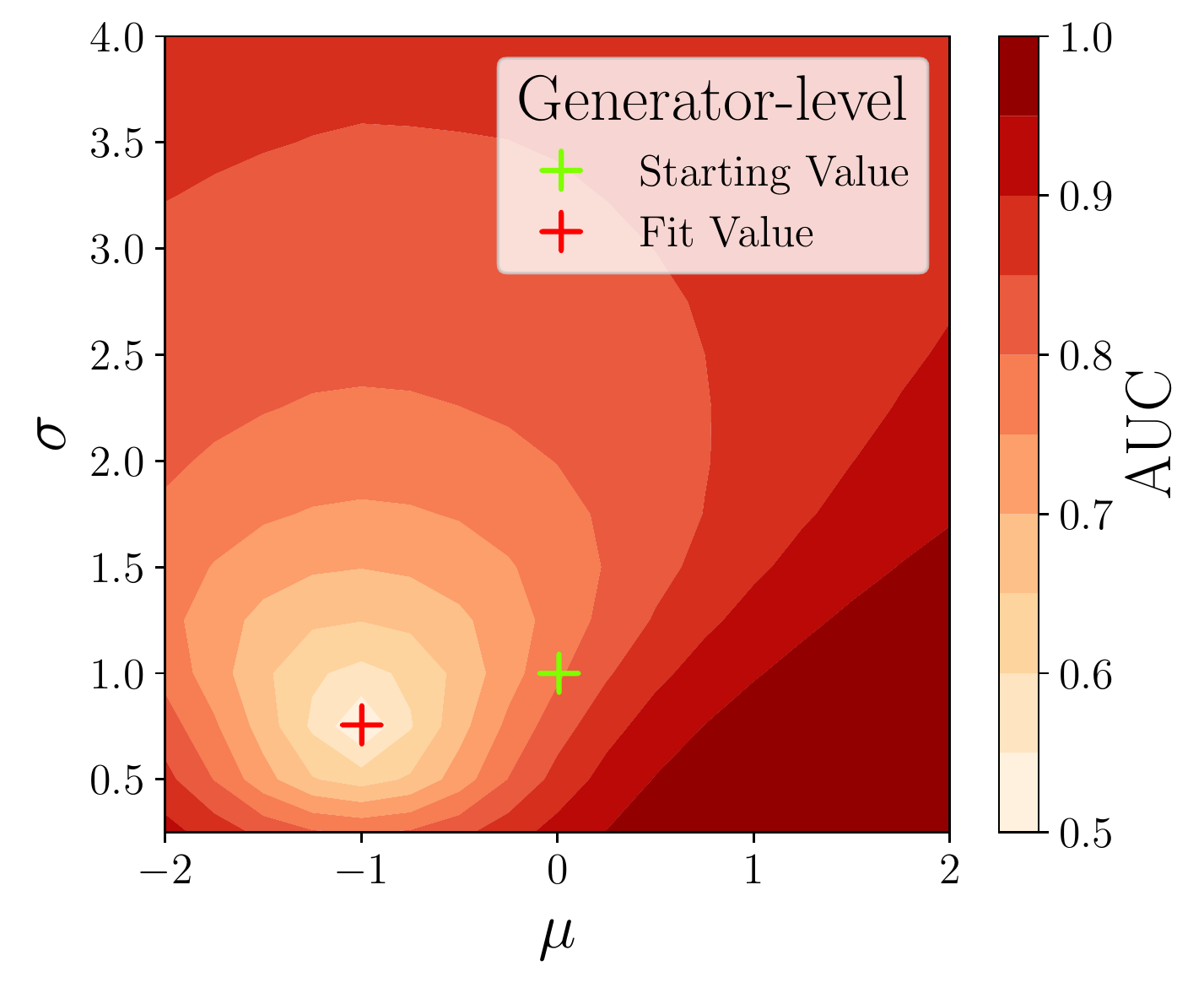}\\
	  \includegraphics[scale=0.5]{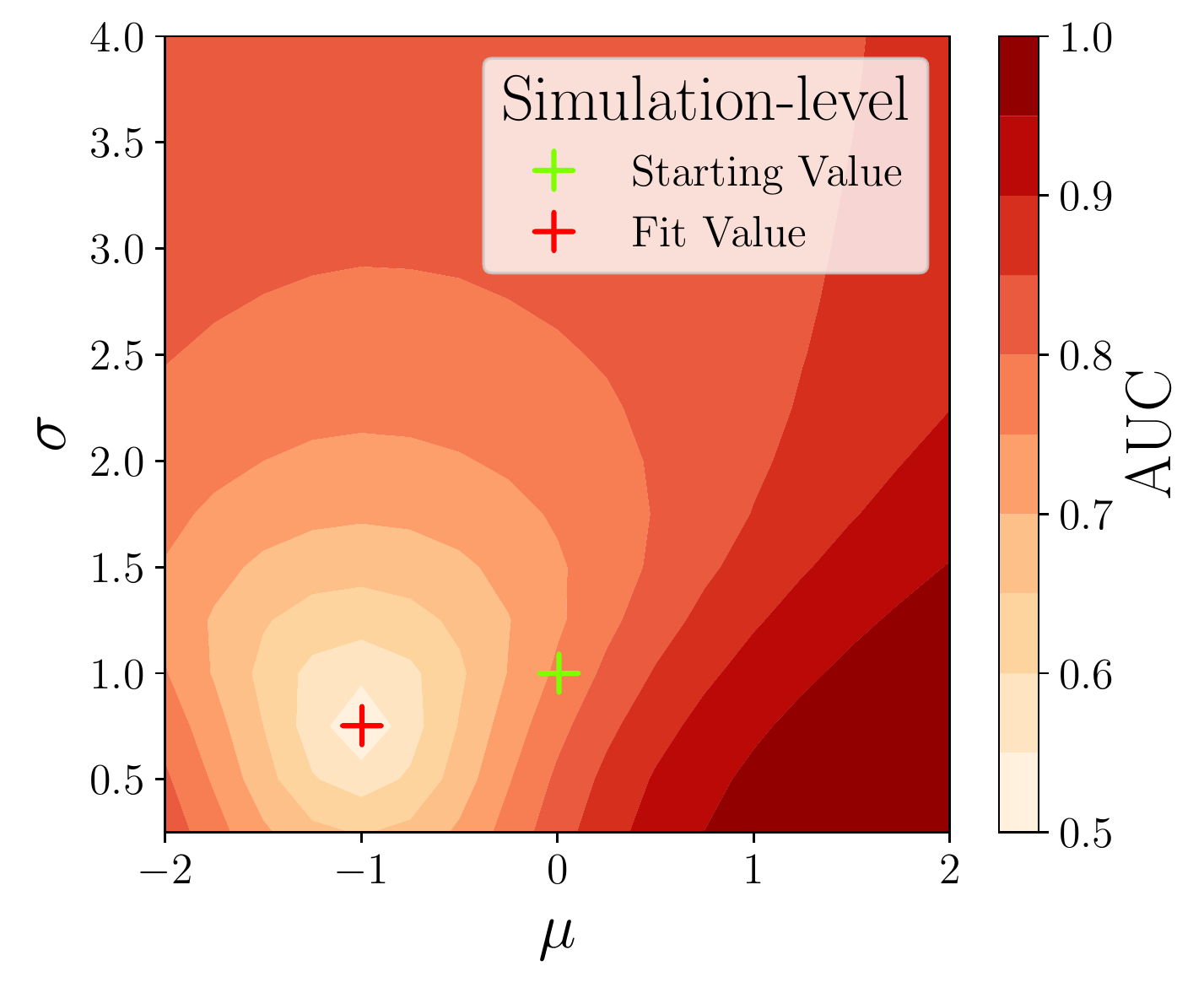}
	  	     \caption{For an individual run, the AUC as a function of $\mu$ and $\sigma$.  The true values are $\mu=-1.000$ and $\sigma=0.750$.  The values of the nominal synthetic dataset are indicated by a green cross and the nondifferentiable optimizer's fitted values are represented by a red cross.}
        \label{fig:gaussian2dauc}
 \end{figure}

A two-dimensional fit to $\mu$ and $\sigma$ is demonstrated in Fig.~\ref{fig:gaussian2dauc}.  The AUC function is minimized at the correct values of $\mu=-1$ and $\sigma=0.75$ for both generator level and simulation level for a reweighting function derived at generator level in both cases.  The contours in Fig.~\ref{fig:gaussian2dauc} indicate that the AUC function rises more steeply away from the minimum at generator level as would be expected of the enhanced statistical power of the dataset without detector effects. The numerical results of the two-dimensional Gaussian fit are presented in Table~\ref{tab:gaussian2dfit}.  The reported measurements and uncertainties are again the averages and standard deviations over 40 runs using a nondifferentiable optimizer. The same procedure for identifying and removing outliers as the one-dimensional fit was employed.

\begin{table}[h!]
	\begin{tabular}{|c|c|c|c|} 
     \hline
     Parameter & Target & Level & Fit value\\ 
     \hline
     \hline
     $\mu$ & -1.000 & \multirow{2}{*}{Generator} & -0.994 $\pm$ 0.014\\
     $\sigma$ & 0.750 & & 0.746 $\pm$ 0.013\\
     \hline
     $\mu$ & -1.000 & \multirow{2}{*}{Simulation} & -0.997 $\pm$ 0.017\\
     $\sigma$ & 0.750 & & 0.747 $\pm$ 0.017\\
     \hline
    \end{tabular}
    \caption{Numerical results for the two-dimensional Gaussian fit.  The reported values and errors represent the mean and standard deviation over 40 runs (with outliers removed), employing the same nondifferentiable optimization scheme described in Fig~\ref{fig:gaussian1dauc} with the addition of fitting $\sigma$ within the bounds [0.25, 4].}
        \label{tab:gaussian2dfit}
\end{table}

\subsection{Parton Shower Monte Carlo Tuning}

The parton shower tuning examples from Ref.~\cite{Andreassen:2019nnm} are presented in this section.  There are no detector effects, but we show that the new fitting methodology works with high-dimensional features and in particular can be integrated with particle flow networks~\cite{Komiske:2018cqr} which are based on deep sets~\cite{10.5555/3294996.3295098}.  The event generator details can be found in Ref.~\cite{Andreassen:2019nnm} and are briefly reviewed here.  In particular, $e^+e^-\rightarrow Z\rightarrow\text{dijets}$ are generated using \textsc{Pythia} 8.230~\cite{Sjostrand:2006za,Sjostrand:2014zea} and anti-$k_t$~\cite{Cacciari:2008gp} $R=0.8$ jets are clustered using \textsc{FastJet} 3.03~\cite{Cacciari:2011ma,Cacciari:2005hq}.  The jets are presented to the neural network for training, with each jet constituent represented by $(p_T, \eta, \phi, \text{particle type}, \theta)$, where $\theta$ are the generator parameters to be determined.  The neural network setup is the same as in Ref.~\cite{Andreassen:2019nnm}, which uses the default particle flow network parameters from Ref.~\cite{Komiske:2018cqr}.  Training time was about 20 seconds per epoch on an NVIDIA Tesla V100 GPU.

The default generator parameters follow the Monash tune~\cite{Skands:2014pea}.  Three representative generator parameters are used here to illustrate the \textsc{Srgn} fitting procedure.  First, \texttt{TimeShower:alphaSvalue} is varied to illustrate a parameter that has a significant impact on the entire phase space and is thus relatively easy to tune.  Second, \texttt{StringZ:aLund} is a parameter that also impacts the entire phase space but to a lesser extent than the strong coupling constant used in final state radiation.  Finally, \texttt{StringFlav:probStoUD} is a parameter that has a large impact on a narrow region of phase space.  The Monash tune values of the three parameters are 0.1365, 0.68, and 0.217, respectively.  For \texttt{TimeShower:alphaSvalue} and \texttt{StringFlav:probStoUD}, two nearly sufficient one-dimensional statistics are known: the number of particles inside the jets and the number of strange hadrons, respectively.  Fits using these simple observables will be compared with the full phase space fit below.

\begin{figure*}[h!]
     \includegraphics[scale=0.4]{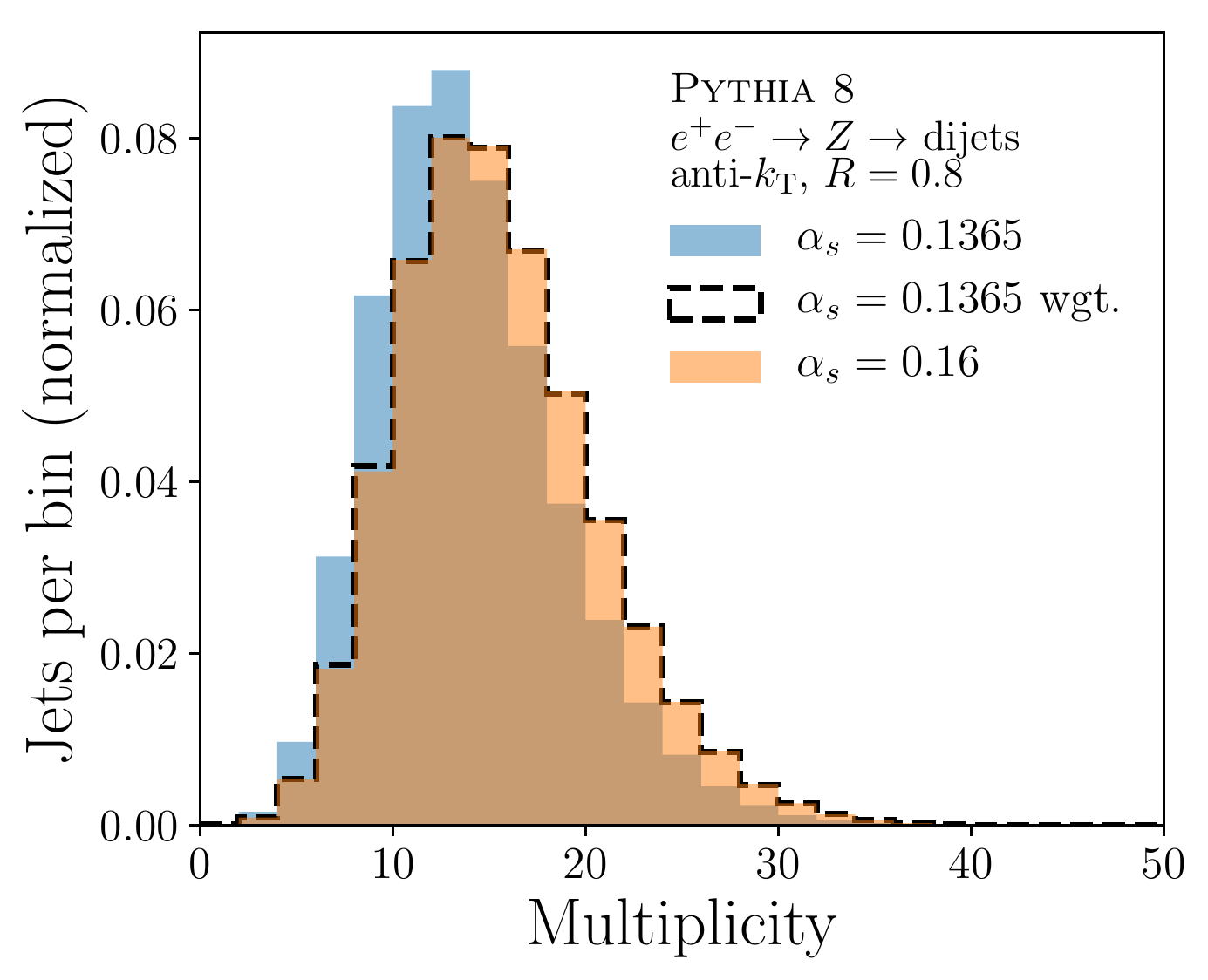}
     \includegraphics[scale=0.4]{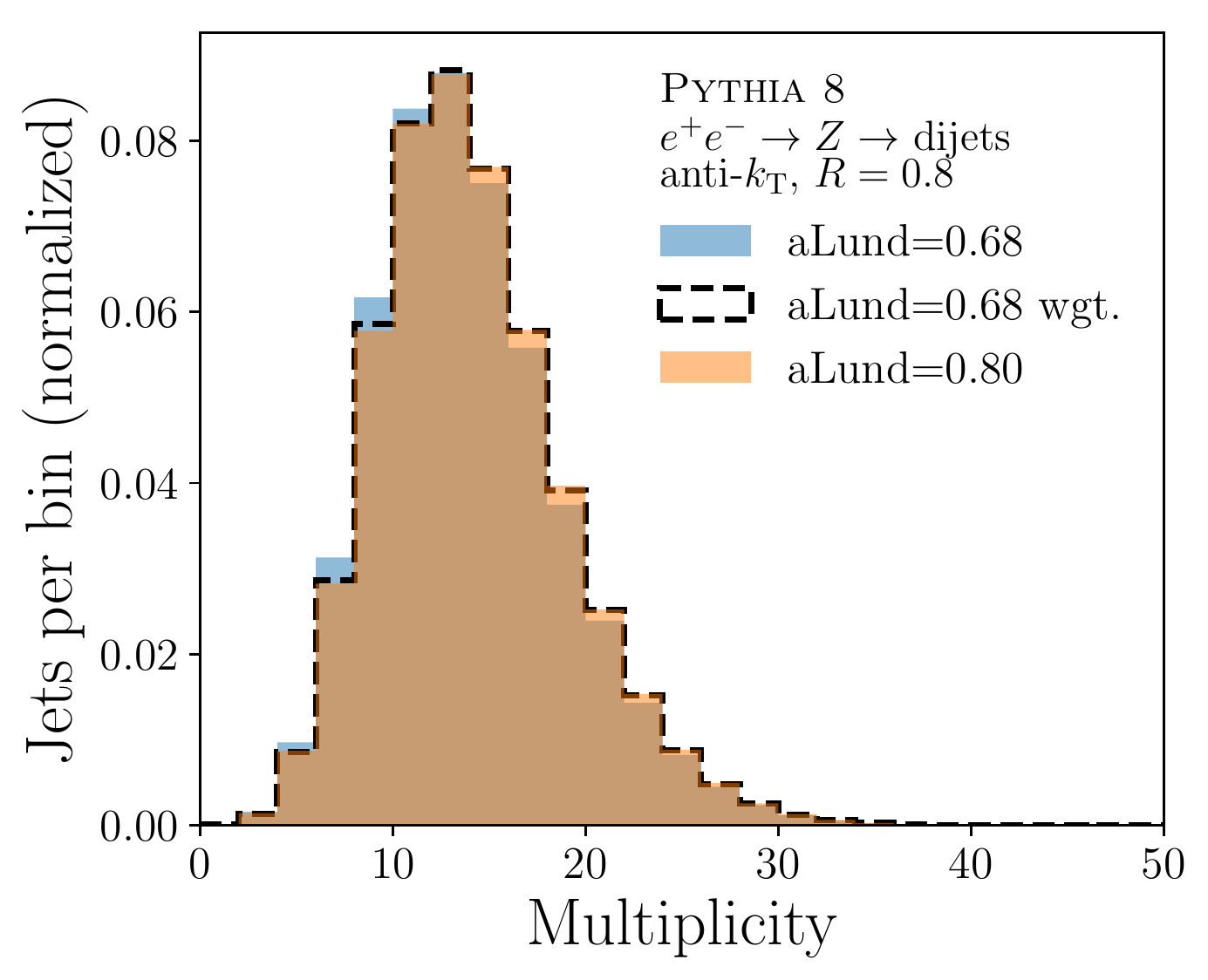}
     \includegraphics[scale=0.4]{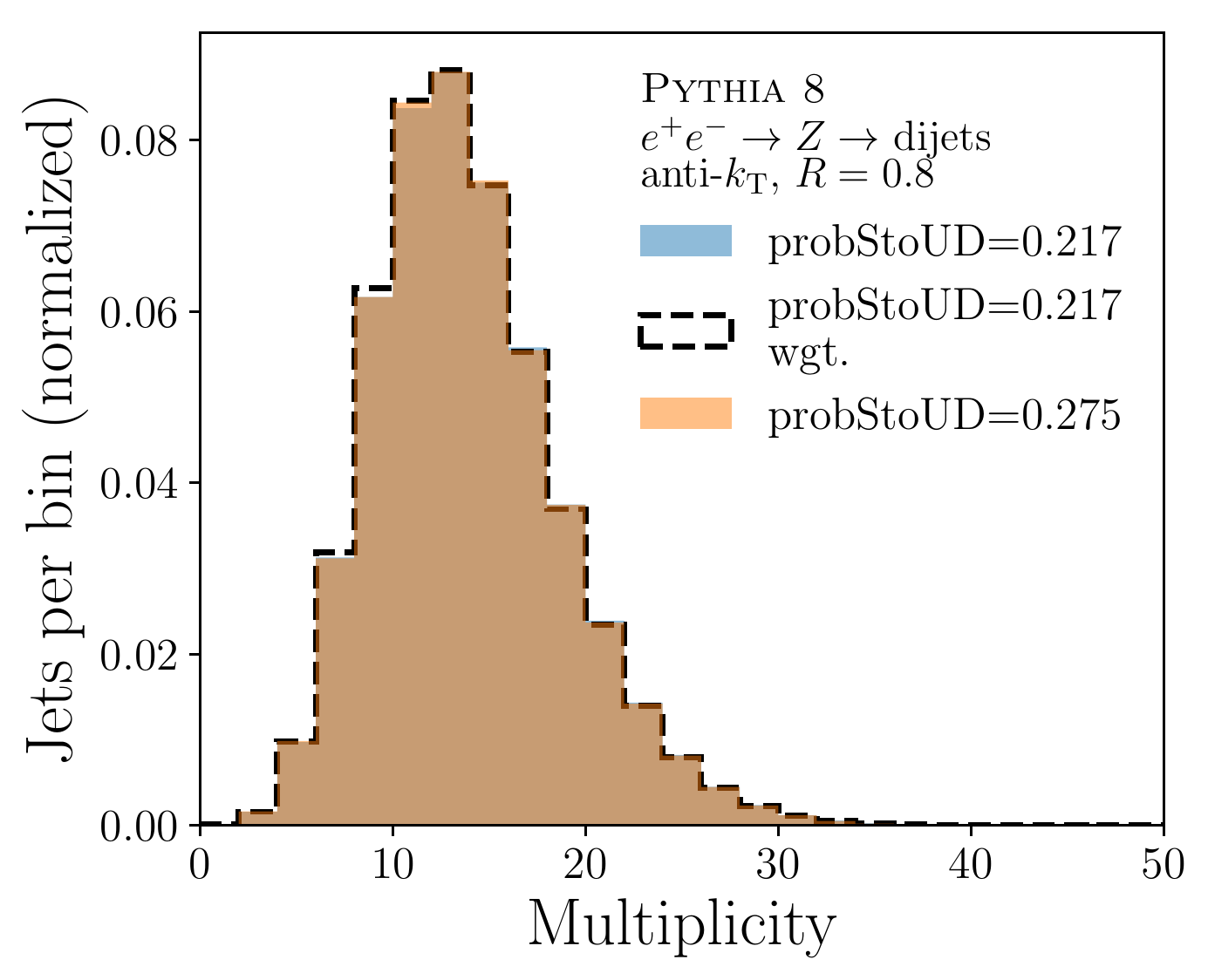}\\
	 \includegraphics[scale=0.4]{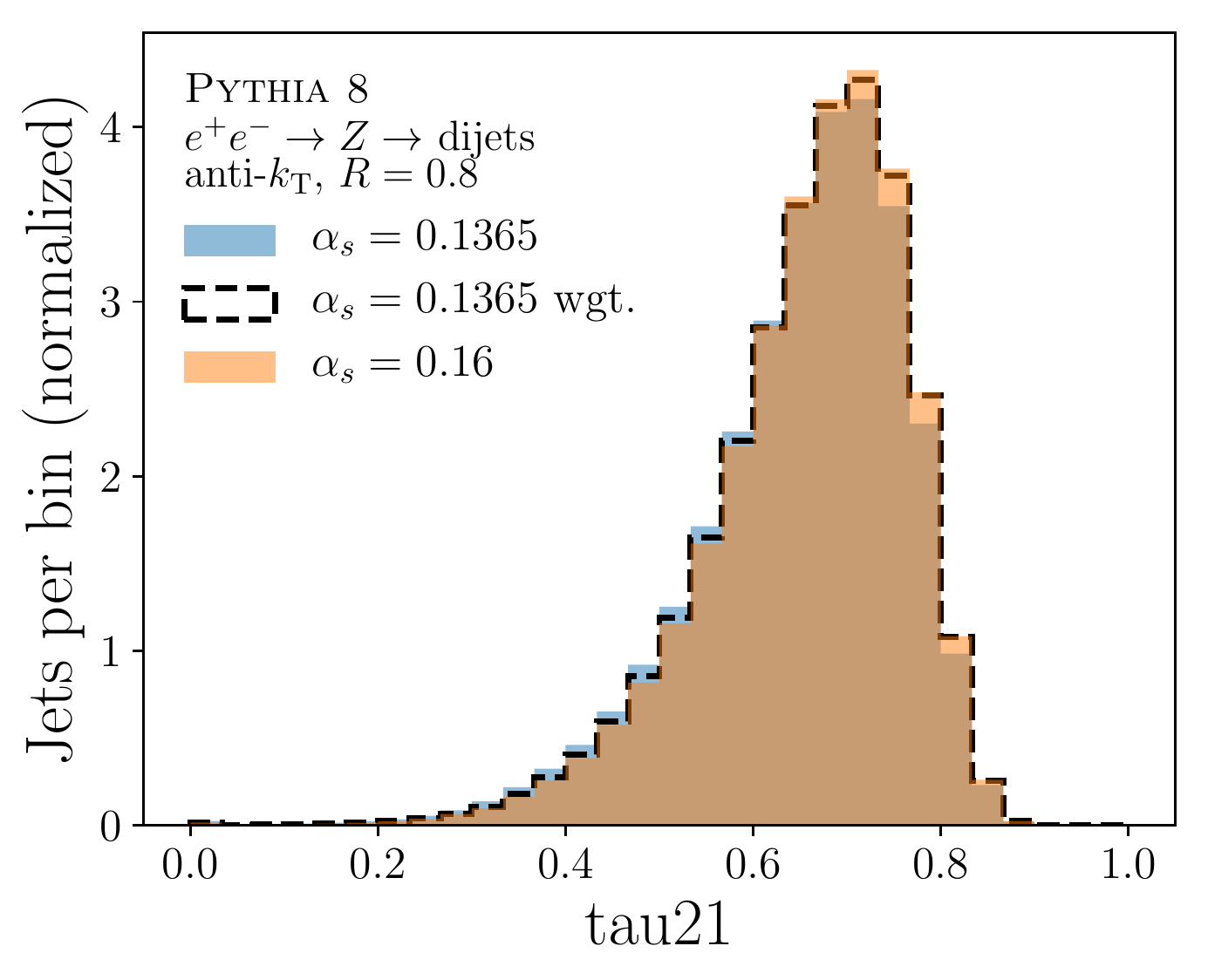}
	  \includegraphics[scale=0.4]{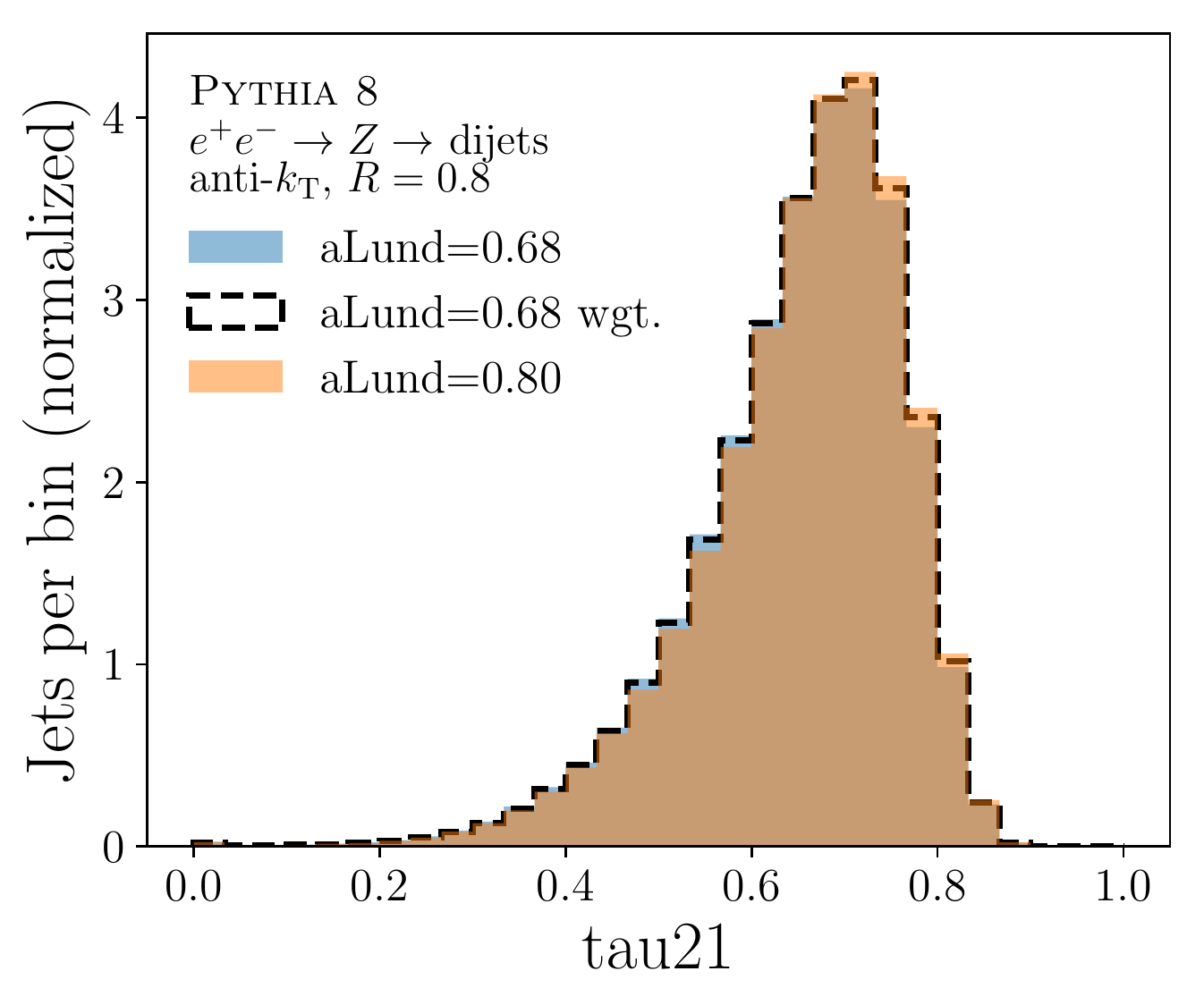}
	   \includegraphics[scale=0.4]{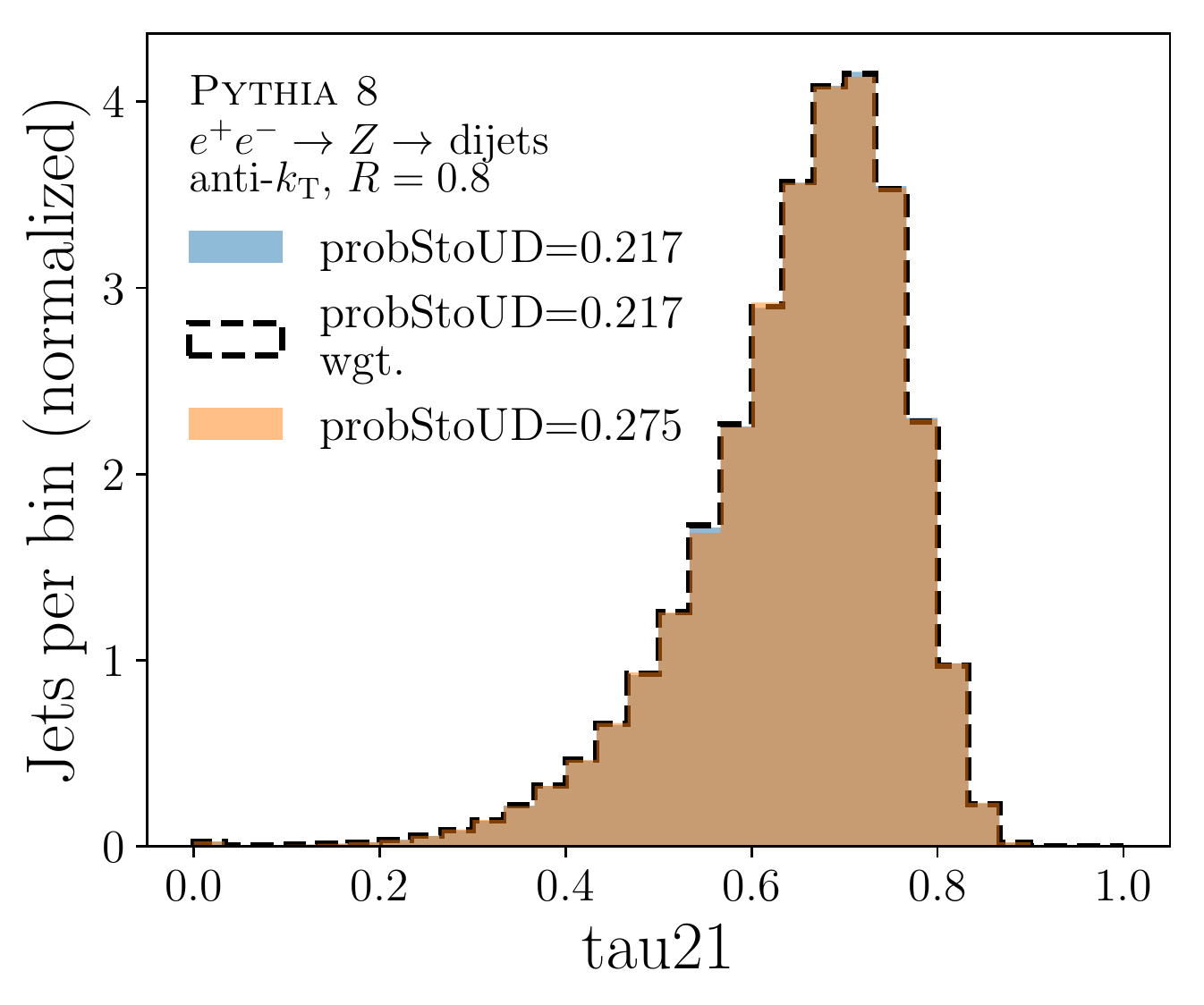}\\
	 \includegraphics[scale=0.4]{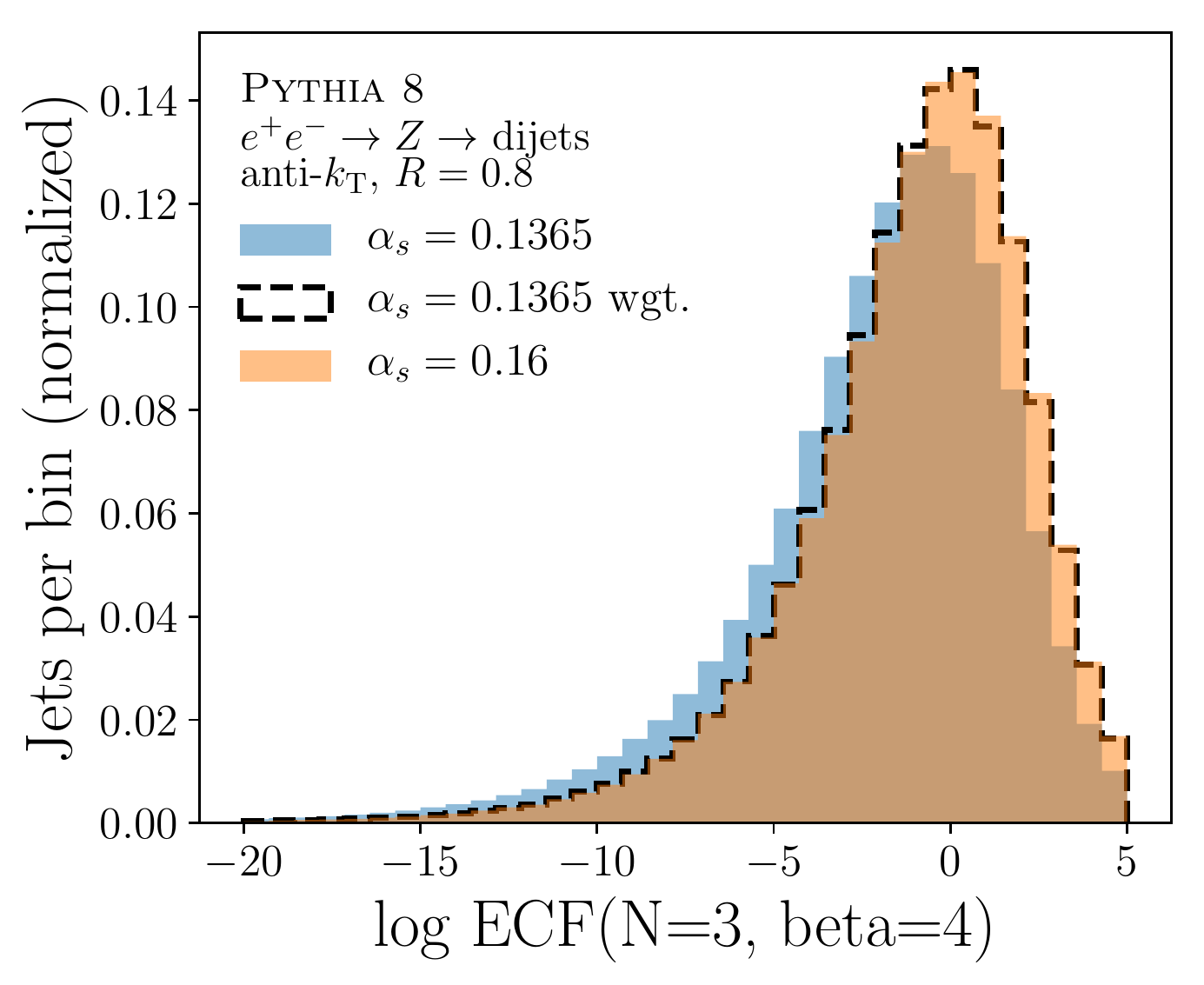}
	 \includegraphics[scale=0.4]{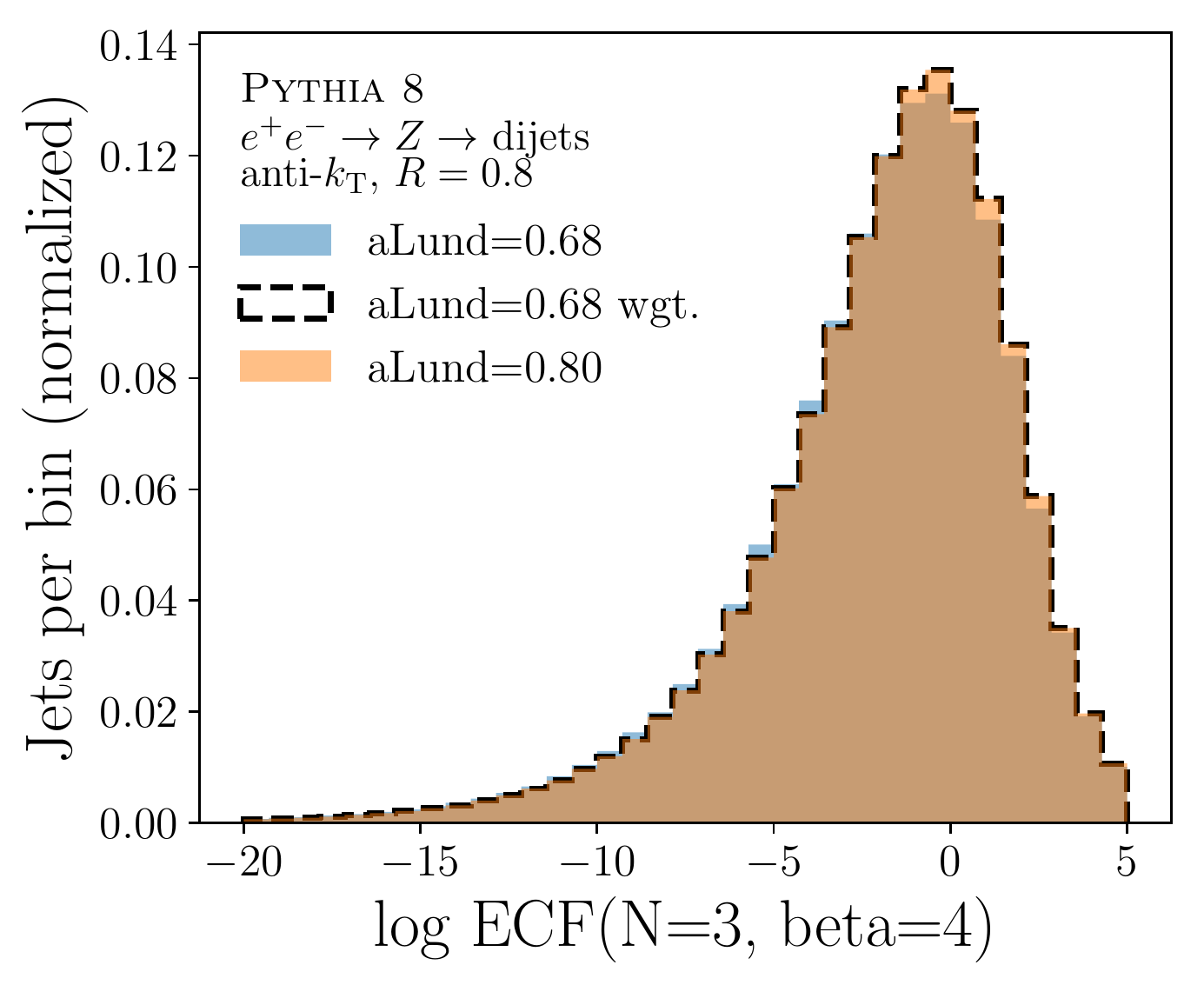}
	  \includegraphics[scale=0.4]{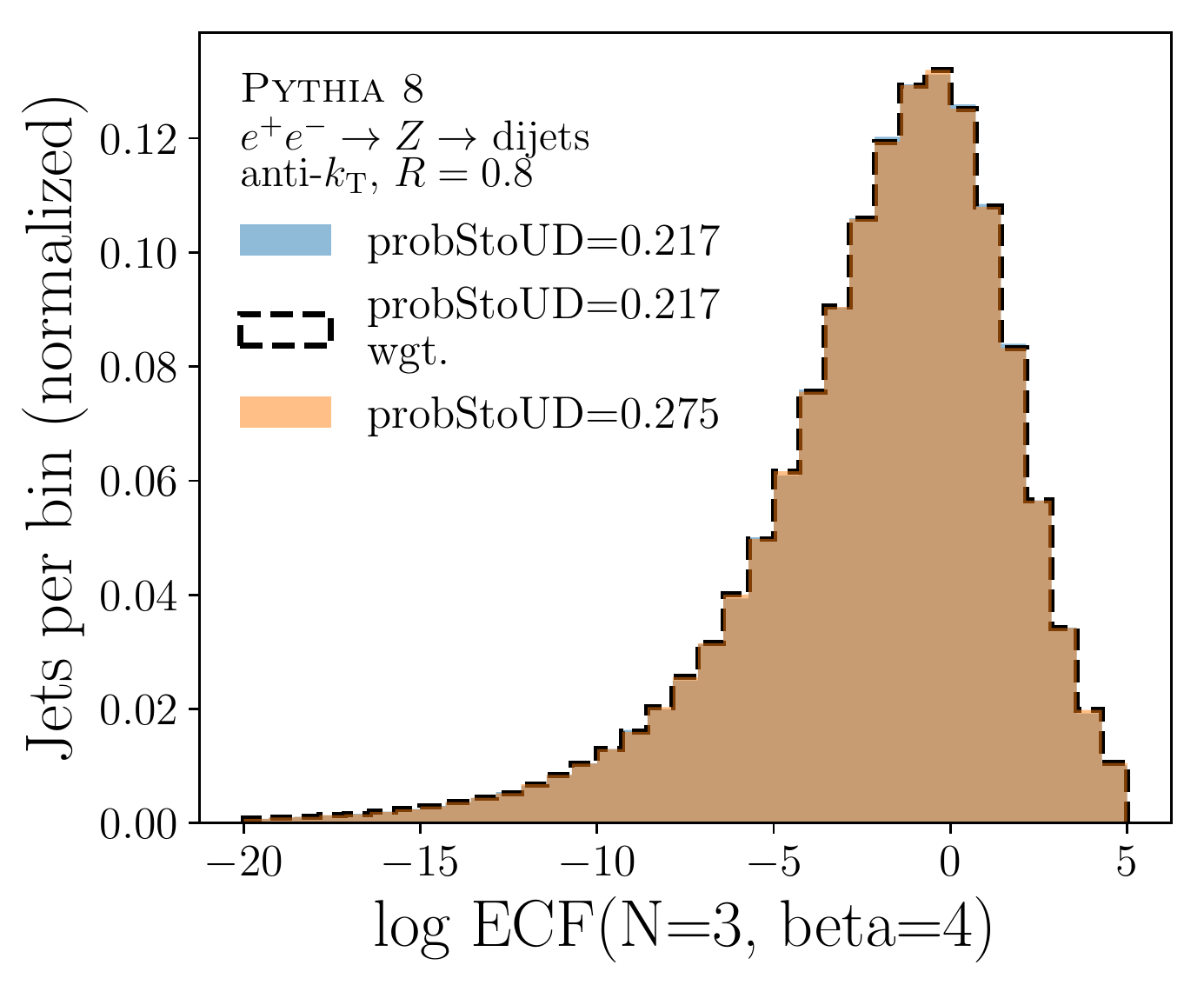}\\
	 \includegraphics[scale=0.4]{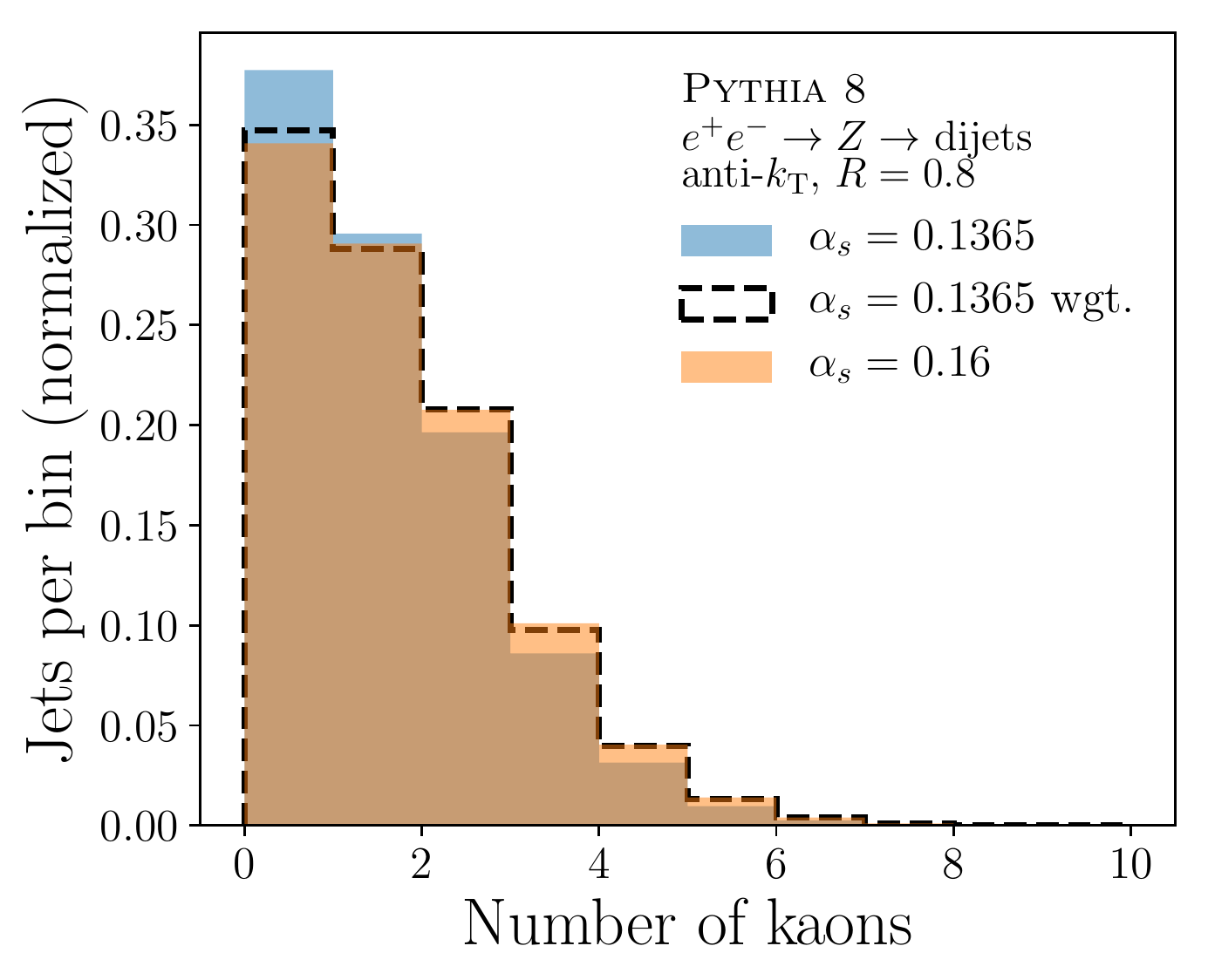} 
	 \includegraphics[scale=0.4]{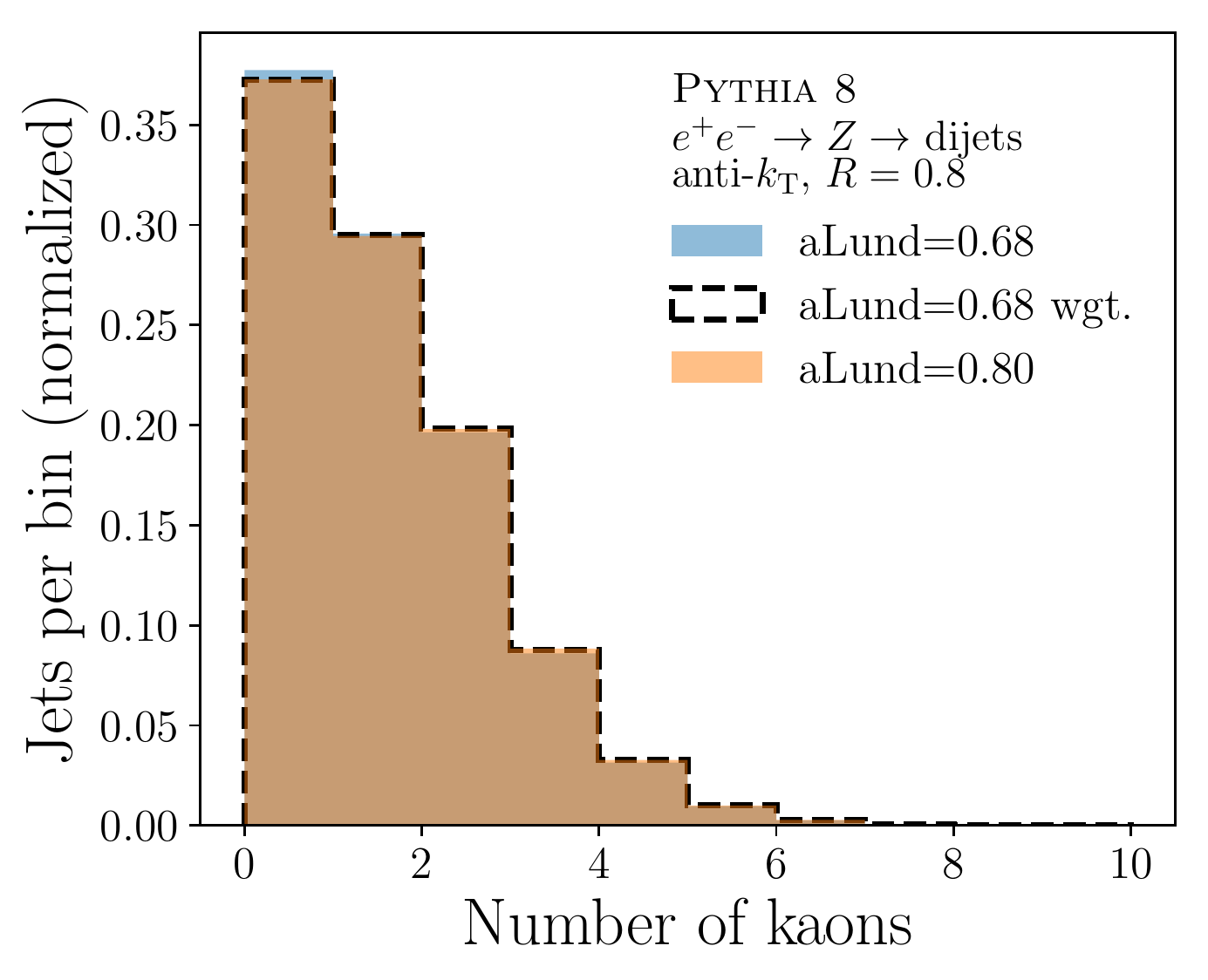}
	 \includegraphics[scale=0.4]{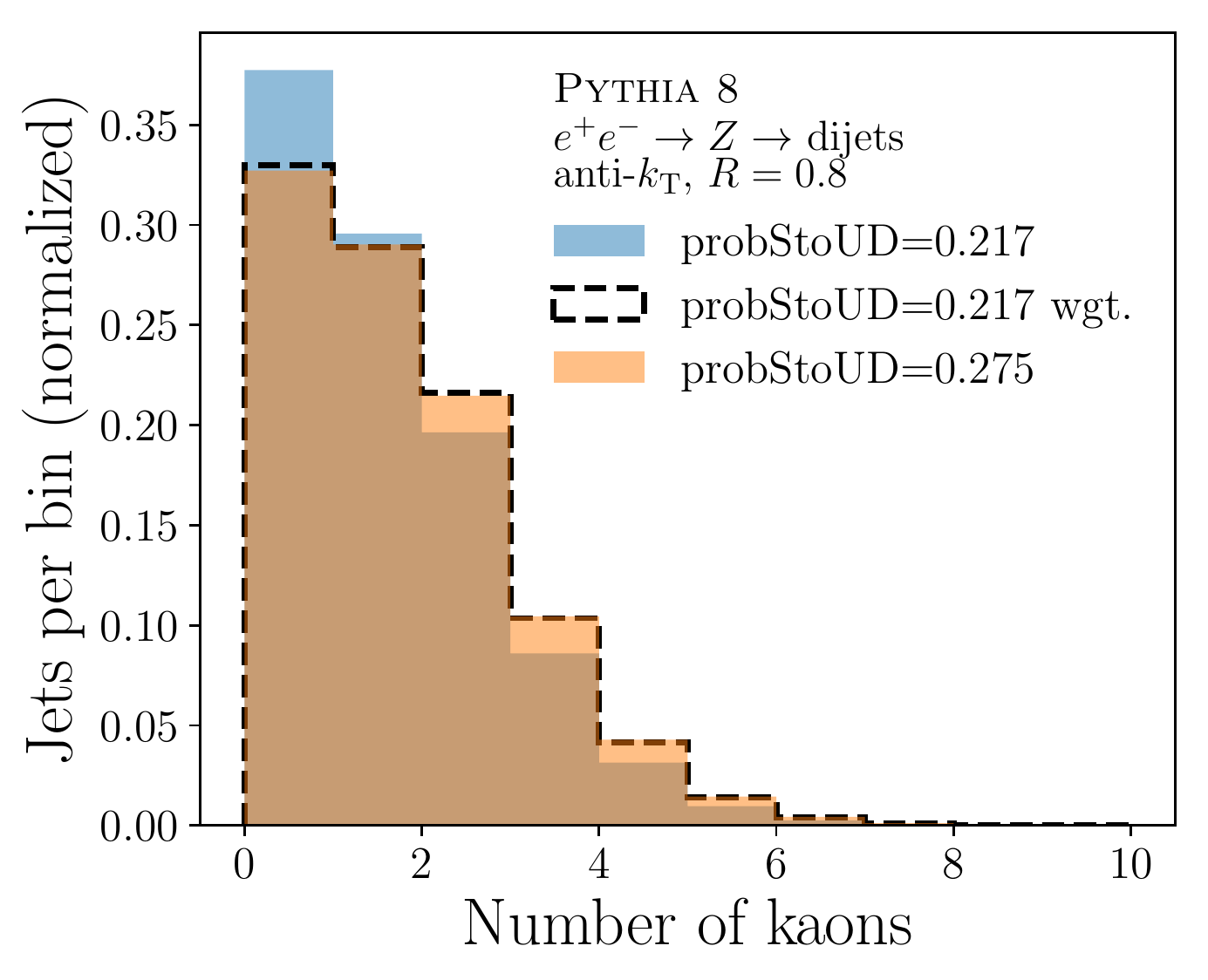}
	    \caption{Features used to show the impact of generator parameter variations for the parton shower dataset.  Variations in \texttt{TimeShower:alphaSvalue}, \texttt{StringZ:aLund}, and \texttt{StringFlav:probStoUD} are presented in the first, second and third columns, respectively.  Each row represents a different observable. Reweighted distributions are plotted over an average of 40 reweightings.}
        \label{fig:partonshowerdataset}
\end{figure*}

Generator-level features illustrating variations in each of the three parameters are shown in Fig.~\ref{fig:partonshowerdataset}.  The full phase space will be used in the fit, but these are representative features to illustrate the effects of parameter variations.  These features are the same as used in Ref.~\cite{Andreassen:2019nnm} and are the number of particles inside the jet (multiplicity), the number of kaons inside the jet, an $n$-subjettiness ratio $\tau_2/\tau_1$~\cite{Thaler:2010tr,Thaler:2011gf}, and a four-point energy correlation function using angular exponent $\beta=4$~\cite{Larkoski:2013eya} $\text{ECF}(N = 3,\beta = 4)$.  As advertised, the final state shower $\alpha_s$ and hadronization parameters affect all four observables, with a bigger shift from $\alpha_s$.  In contrast, the strangeness parameter only affects the number of kaons and has no impact on the other observables.  To perform a given fit, we scan for the AUC as a function of the parameter to search for the minimum; the step sizes are 0.001, 0.01, and 0.005 for \alphas, \lund, and \strange,~respectively.

\begin{figure}[h!]
  \includegraphics[scale=0.45]{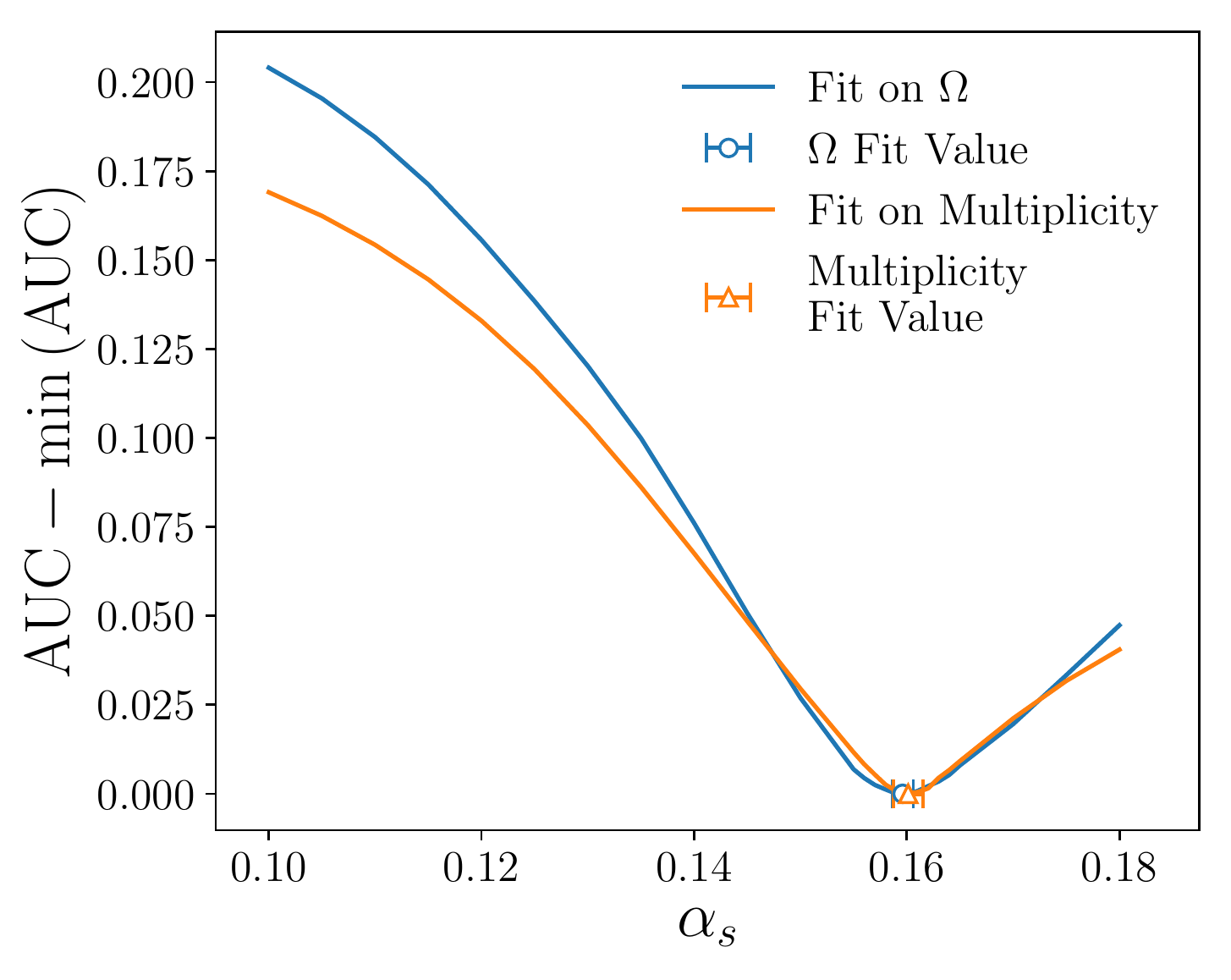}\\
  \includegraphics[scale=0.45]{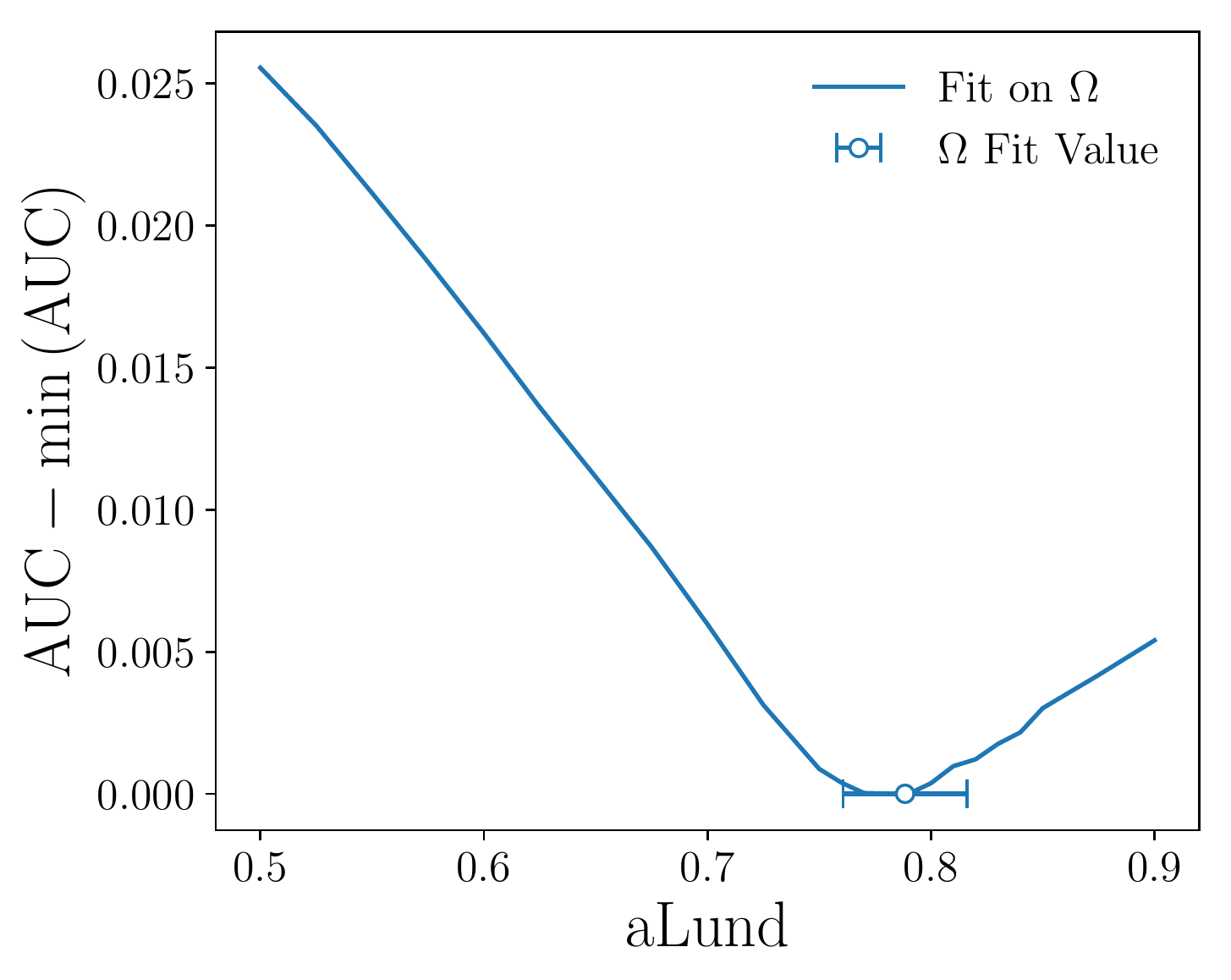}\\
   \includegraphics[scale=0.45]{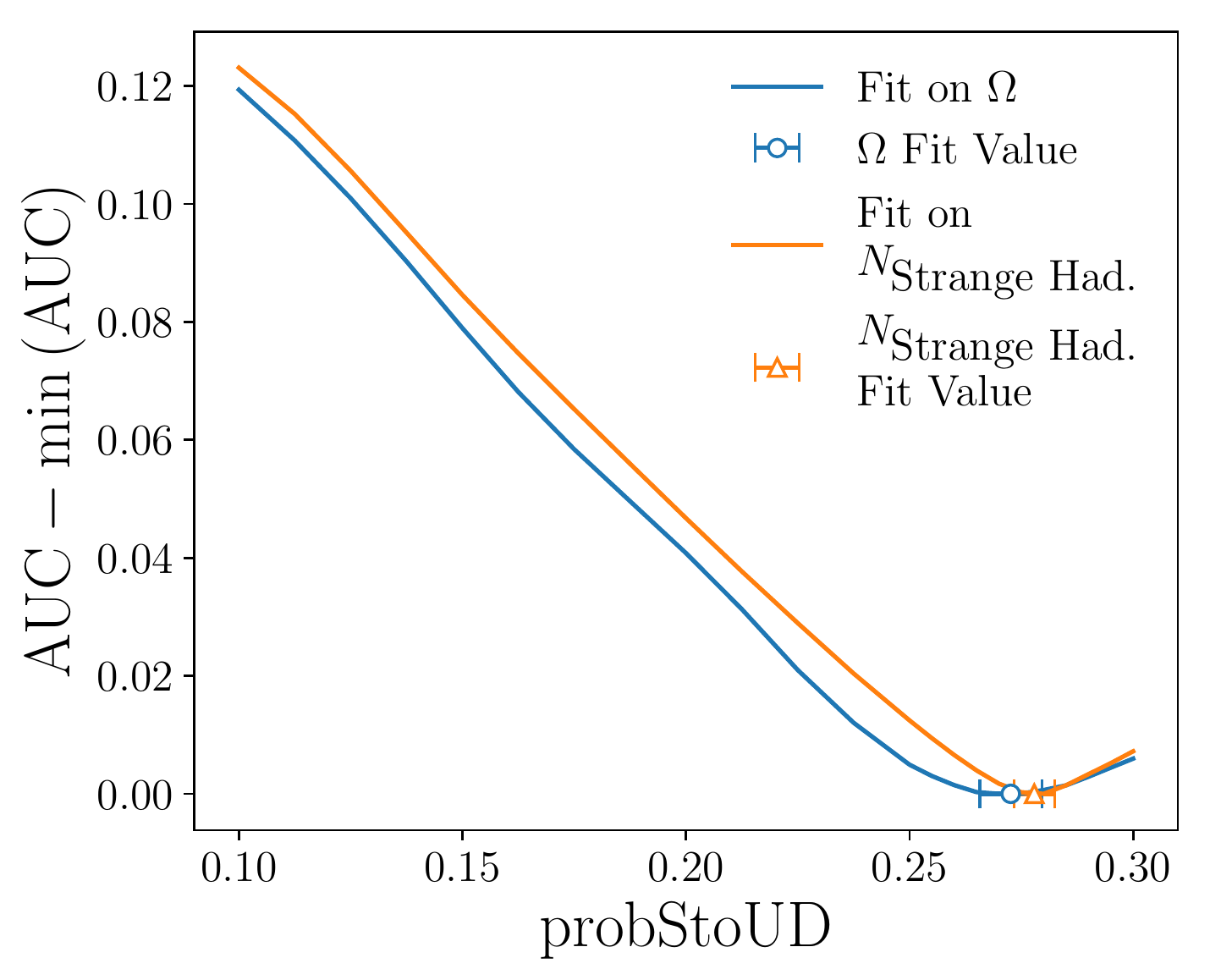}
	    \caption{One-dimensional fits for each of the three parton shower parameters.  The vertical axes show the increase of the AUC for the classifier $g$ from its minimum value.  For comparison, the $\alpha_s$ and strangeness plots also show fits using \textsc{Srgn} with only the inclusive or strange hadron multiplicity, respectively.}
        \label{fig:1dfitspartonshower}
\end{figure}

One-dimensional fits to each of the three parton shower parameters are shown in Fig.~\ref{fig:1dfitspartonshower}.  Since \alphas~has such a large effect on the phase space, it is the most precisely measured parameter as indicated by the steepness of the AUC curve near the minimum.  The steepness of the full phase space fit also shows that there is slightly more information with respect to multiplicity alone.  The \lund~parameter has the smallest effect on the phase space of all three parameters and thus is the least precisely measured parameter.  \strange~primarily has an effect on the number of strange particles, and thus the full phase space does not offer much more information than only the number of strange hadrons, so the precision is comparable for both approaches.  The reported measurements and plots are the averages and standard deviations over 40 runs, each with a different reweighting function and classifier that differed only in their random initialization.  A small number of the runs resulted in reweighting functions that were defective and these were identified and removed by examining the runs with fitted values outside a $2\sigma$ window around the mean.  Across the 40 runs, most of the results clustered around the mean, so the outliers look systematically different than the fits with effective reweighting functions.

%\begin{adjustbox}
	\begin{table}
	\begin{tabular}{|c|c|c|c|} 
     \hline
     Parameter & Target & Input & Fit value \\ 
     \hline
     \hline
     \texttt{TimeShower:} & \multirow{2}{*}{0.1600} & $\Omega$ & 0.1596 $\pm$ 0.0010 \\
     \texttt{alphaSvalue}& & Multiplicity & 0.1601 $\pm$ 0.0014 \\
     \hline
     \lund & 0.8000 & $\Omega$ & 0.7884 $\pm$ 0.0277 \\
     \hline
     \texttt{StringFlav:} & \multirow{2}{*}{0.2750} & $\Omega$ & 0.2726 $\pm$ 0.0070\\
     \texttt{probStoUD}& & $N_{\text{Strange Had.}}$ & 0.2779 $\pm$ 0.0045 \\
     \hline 
    \end{tabular}
    \caption{Numerical results for the parton shower parameter fits.  The errors represent the standard deviation over 40 runs (with outliers removed).  Note that $\Omega$ denotes the full phase space.}
    \label{tab:partonshowerfit}
    \end{table}
%\end{adjustbox}

The numerical results of the three fits are presented in Table~\ref{tab:partonshowerfit}.  The fitted values are statistically consistent with the target values and the uncertainties are generally comparable to or smaller than the values from the original \textsc{Dctr} protocol~\cite{Andreassen:2019nnm}.

\subsection{Top Quark Mass}

Top quark pair production is generated using \textsc{Pythia} 8.230~\cite{Sjostrand:2006za,Sjostrand:2014zea} and detector effects are modeled using \textsc{Delphes} 3.4.1~\cite{deFavereau:2013fsa,Mertens:2015kba,Selvaggi:2014mya} using the default compact muon solenoid (CMS) run card.   One of the $W$ bosons is forced to decay to $\mu+\nu_\mu$ while the other $W$ boson decays hadronically.  Each event is recorded as a variable-length set of objects, consisting of jets, muons, and neutrinos.  At simulation level, the neutrino is replaced with missing transverse momentum.  Generator-level and simulation-level jets are clustered with the anti-$k_t$ algorithm using $R=0.4$ and are labeled as $b$ tagged if the highest energy parton inside the jet cone ($\Delta R < 0.5$) is a $b$ quark.  Jets are required to have $p_T>20$~GeV and they can only be $b$ tagged if $|\eta|<2.5$.  Furthermore, jets overlapping with the muon are removed.

Events are only saved if they have at least two $b$-tagged jets and at least two additional non-$b$-tagged jets.  Four observables are formed for performing the top quark mass extraction.  First, the $b$ jet closest to the muon is labeled $b_1$.  Of the remaining $b$-tagged jets, the highest $p_T$ one is labeled $b_2$.  The highest two\footnote{This has a higher efficiency than imposing a $W$ mass constraint, but there is no significant impact on the results.} $p_T$ non-$b$-tagged jets are labeled $j_1$ and $j_2$.  The four observables are given by: $m_{b_1\mu\nu}$, $m_{b_2\mu\nu}$, $m_{b_1j_1j_2}$, and $m_{b_2j_1j_2}$, where the four-momentum of the detector-level neutrino is determined by solving the quadratic equation for the $W$ boson mass\footnote{If there is no solution, the mass is set to zero.  If there are two real solutions, the one with the smaller $|p_z|$ is selected.}.

\begin{figure*}[h!]
     \includegraphics[scale=0.4]{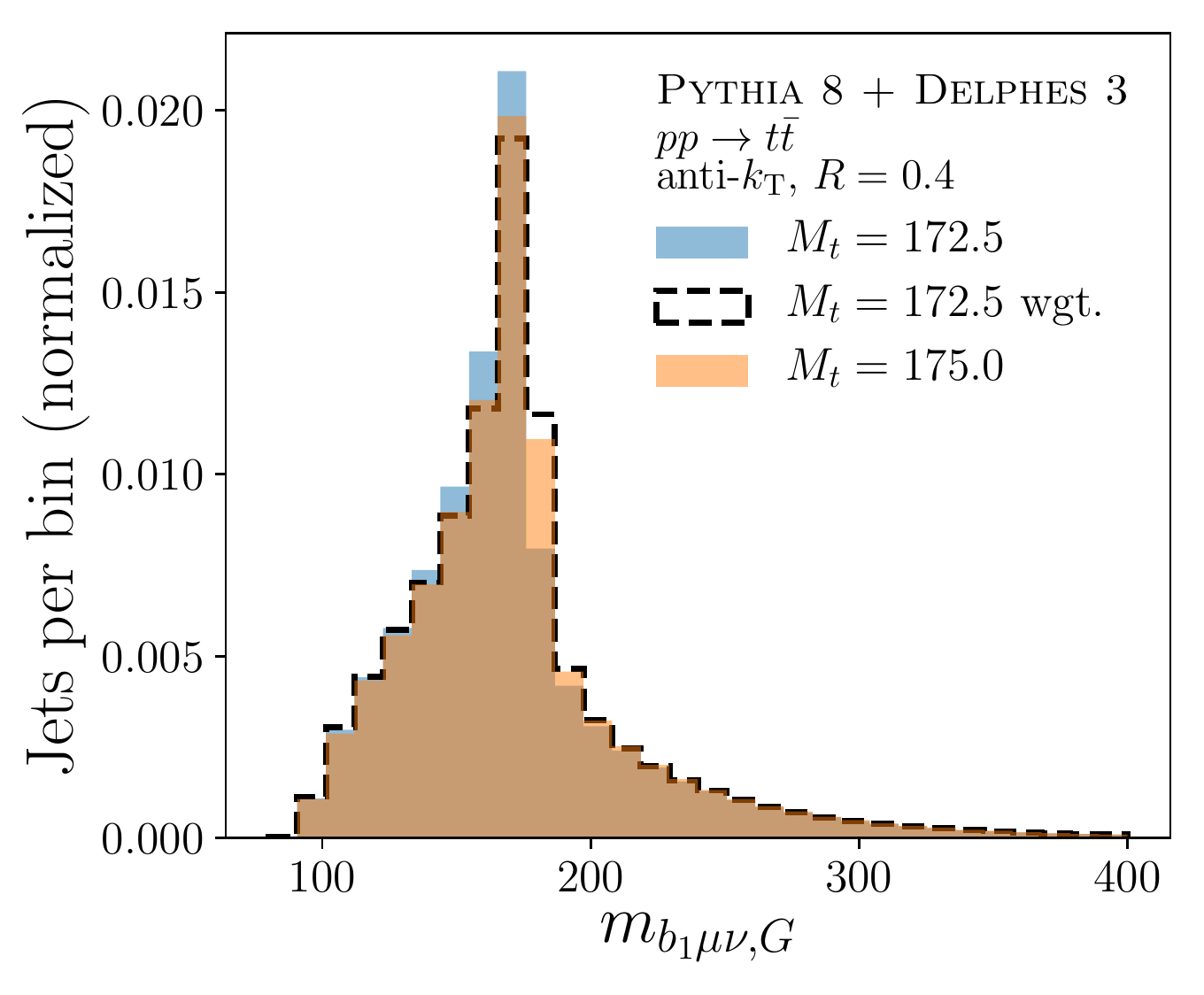}
     \includegraphics[scale=0.4]{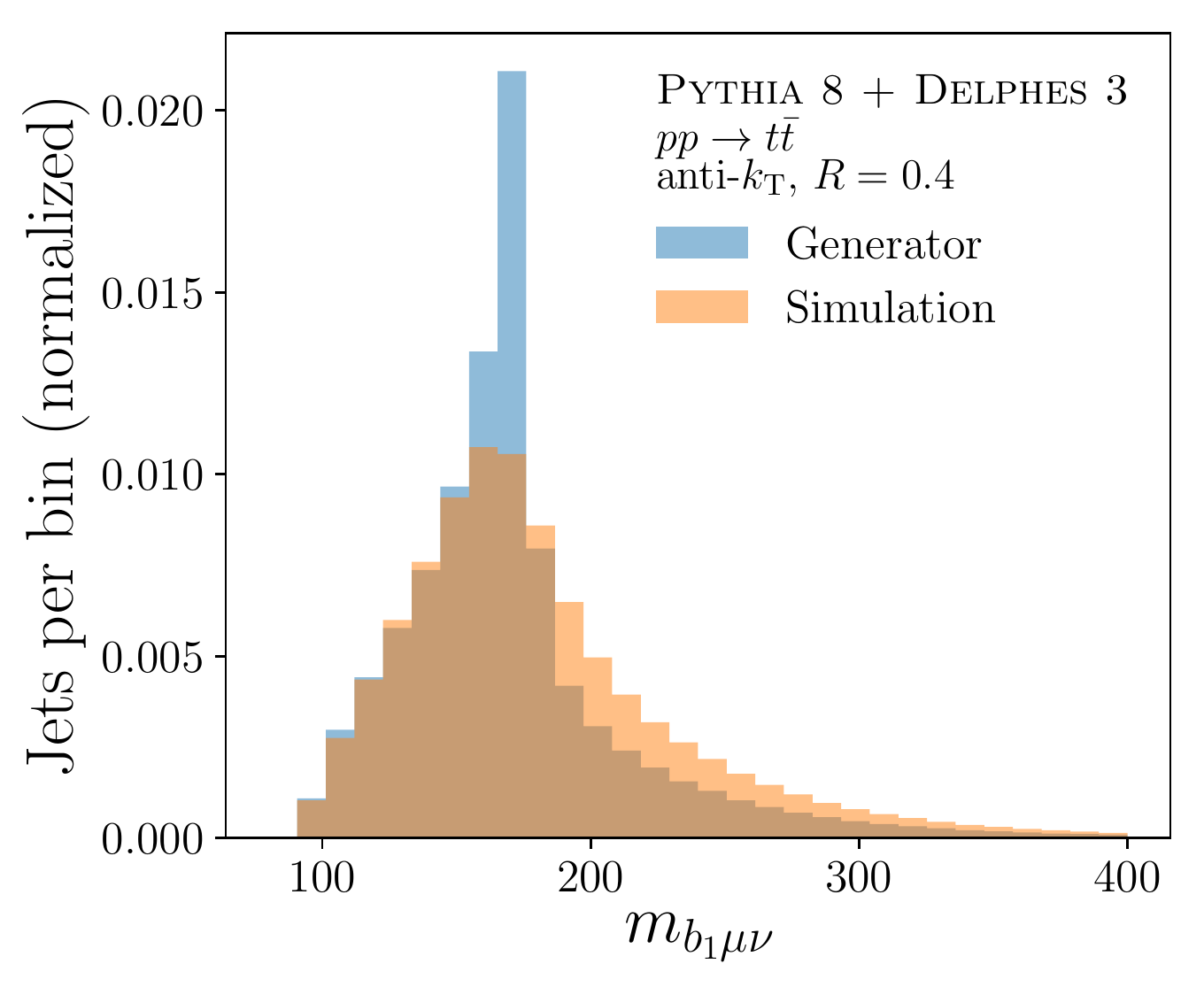}
      \includegraphics[scale=0.4]{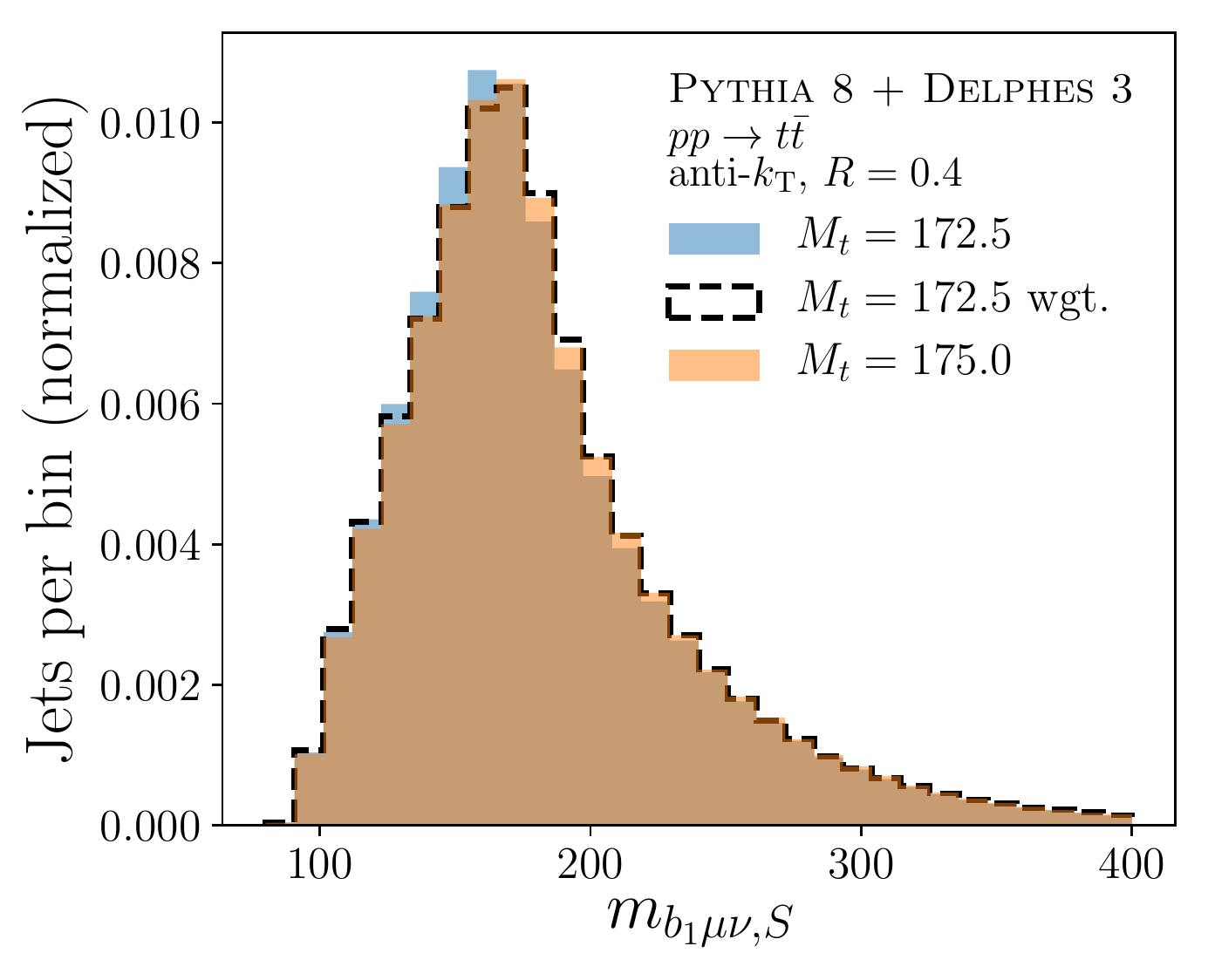}\\
	  \includegraphics[scale=0.4]{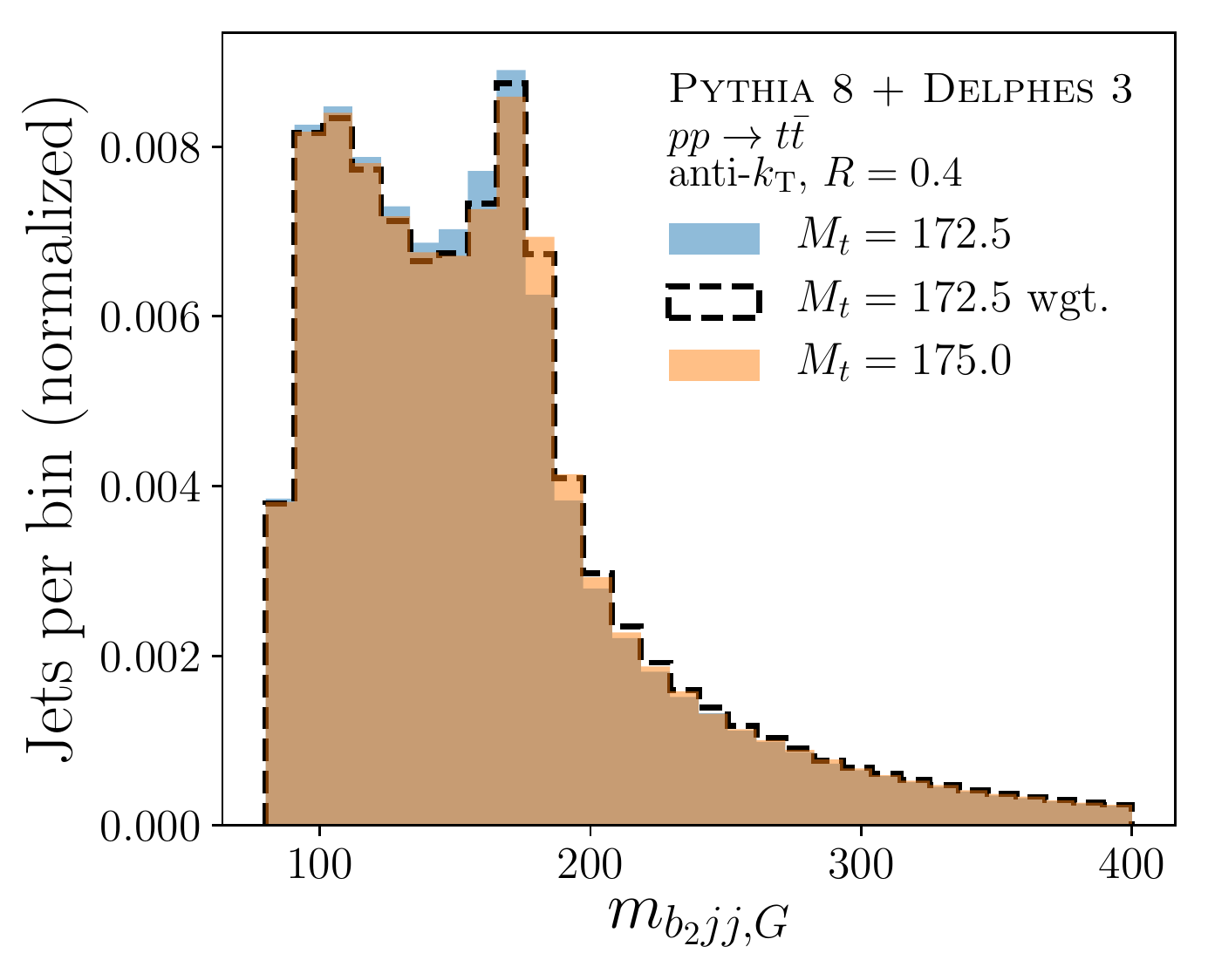}
	  \includegraphics[scale=0.4]{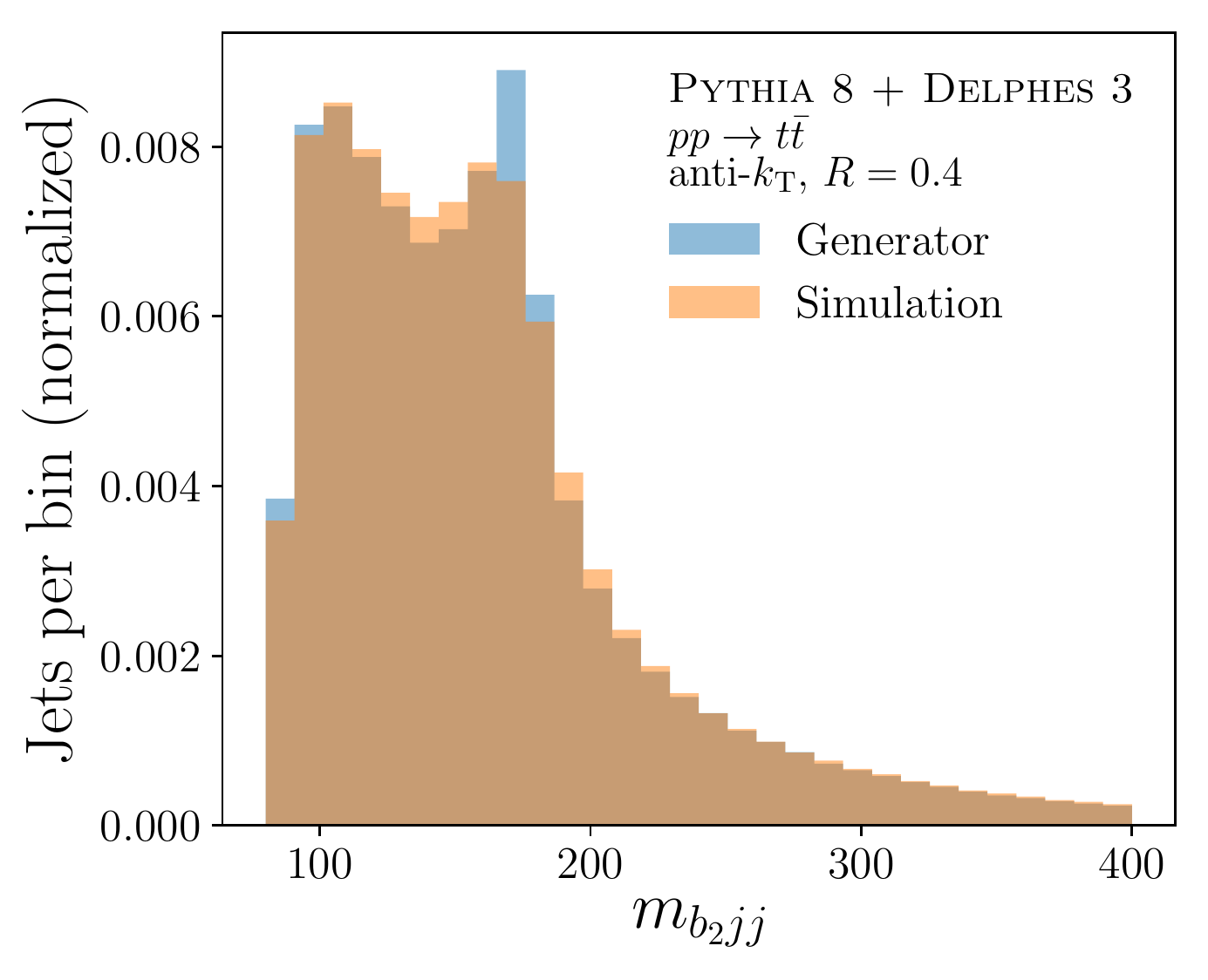}
	  \includegraphics[scale=0.4]{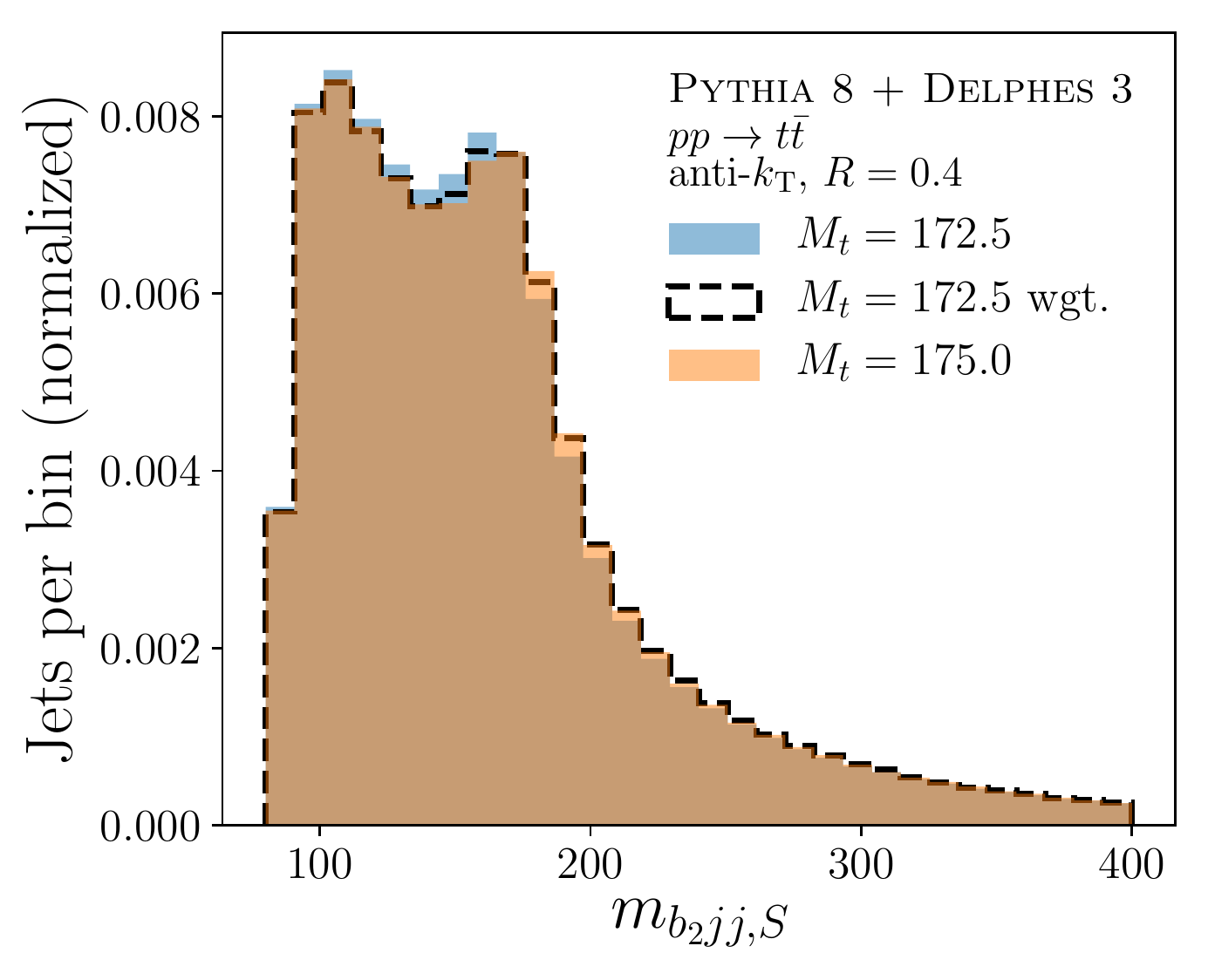}\\
	  
	  \includegraphics[scale=0.4]{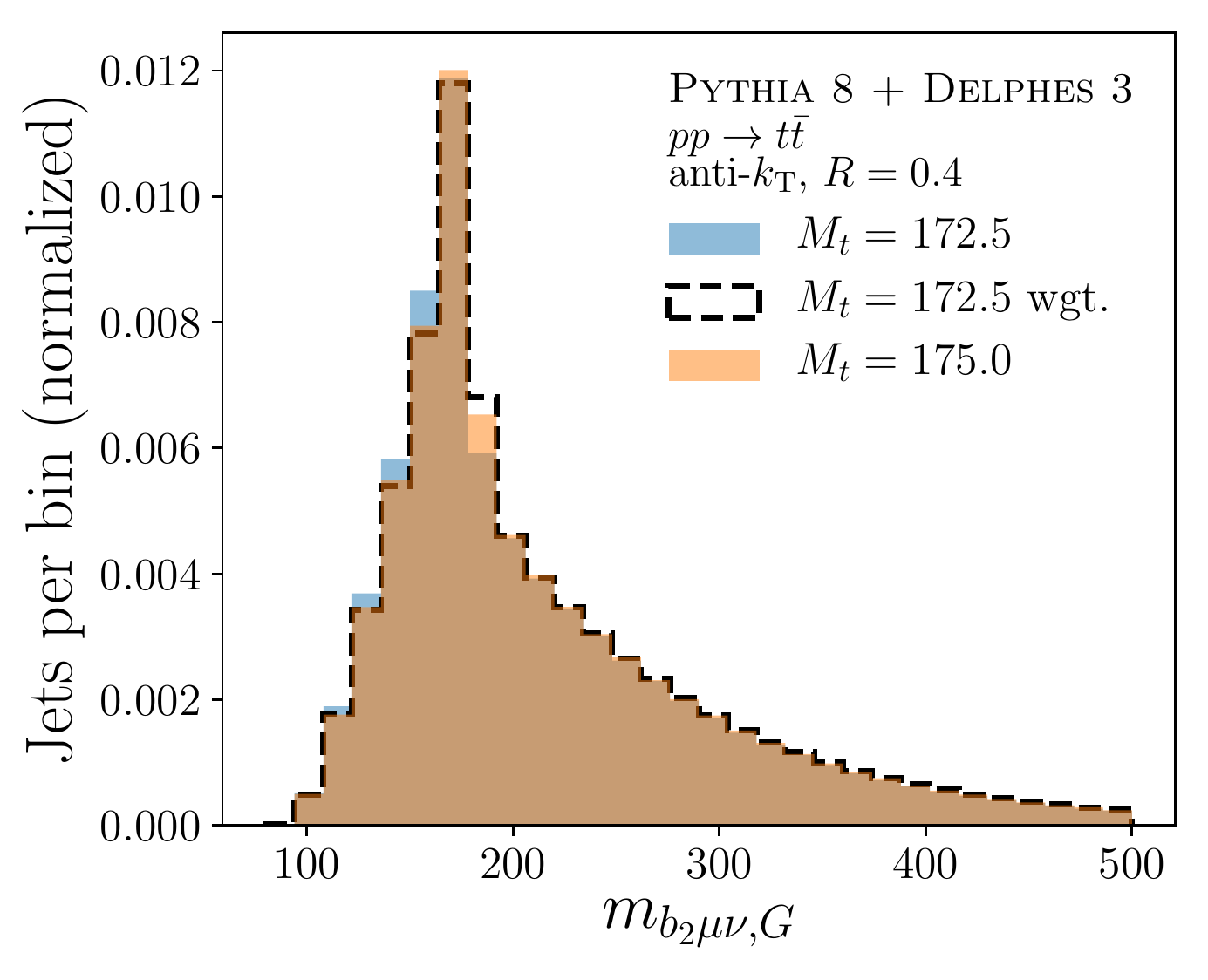}
	  \includegraphics[scale=0.4]{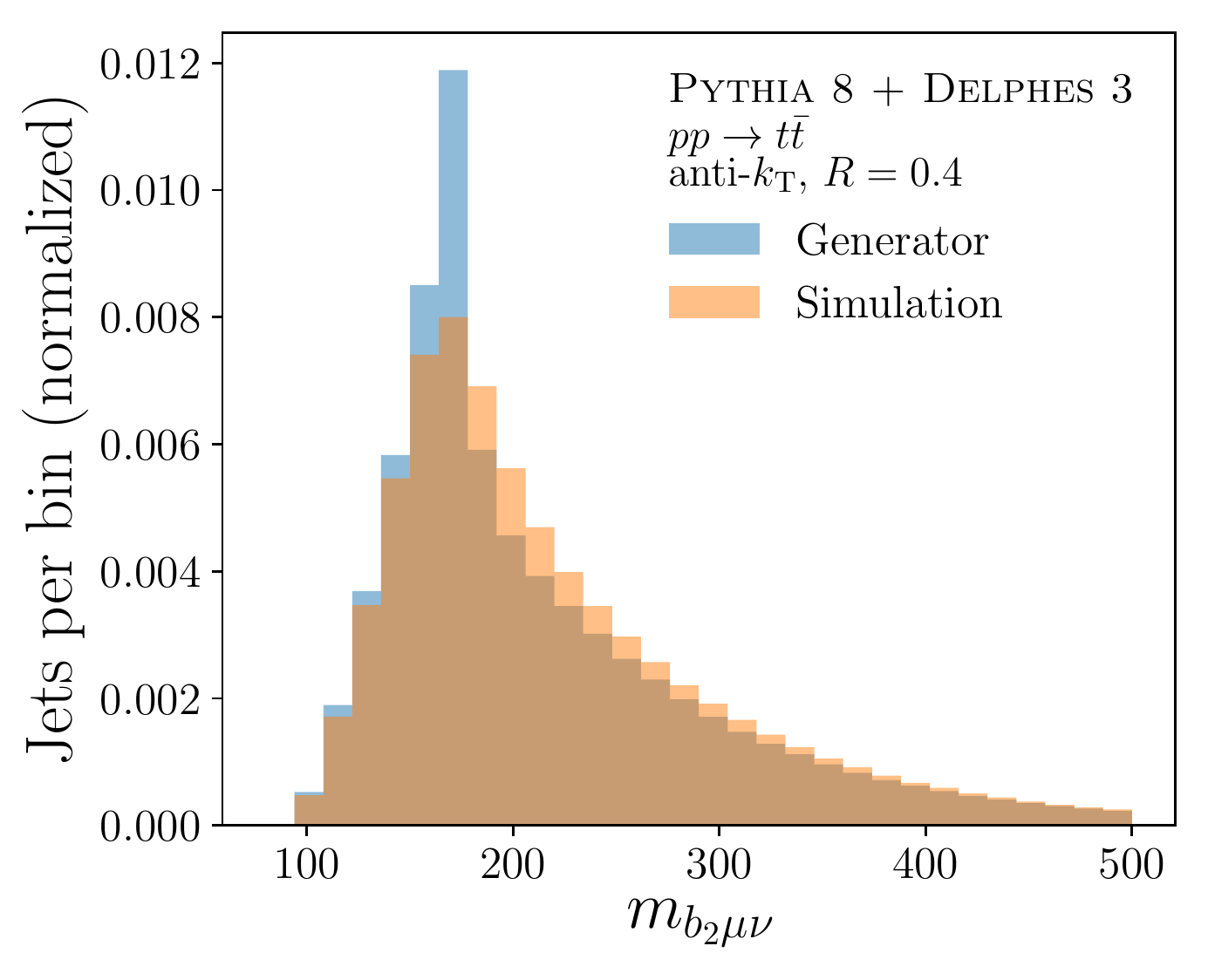}
	  \includegraphics[scale=0.4]{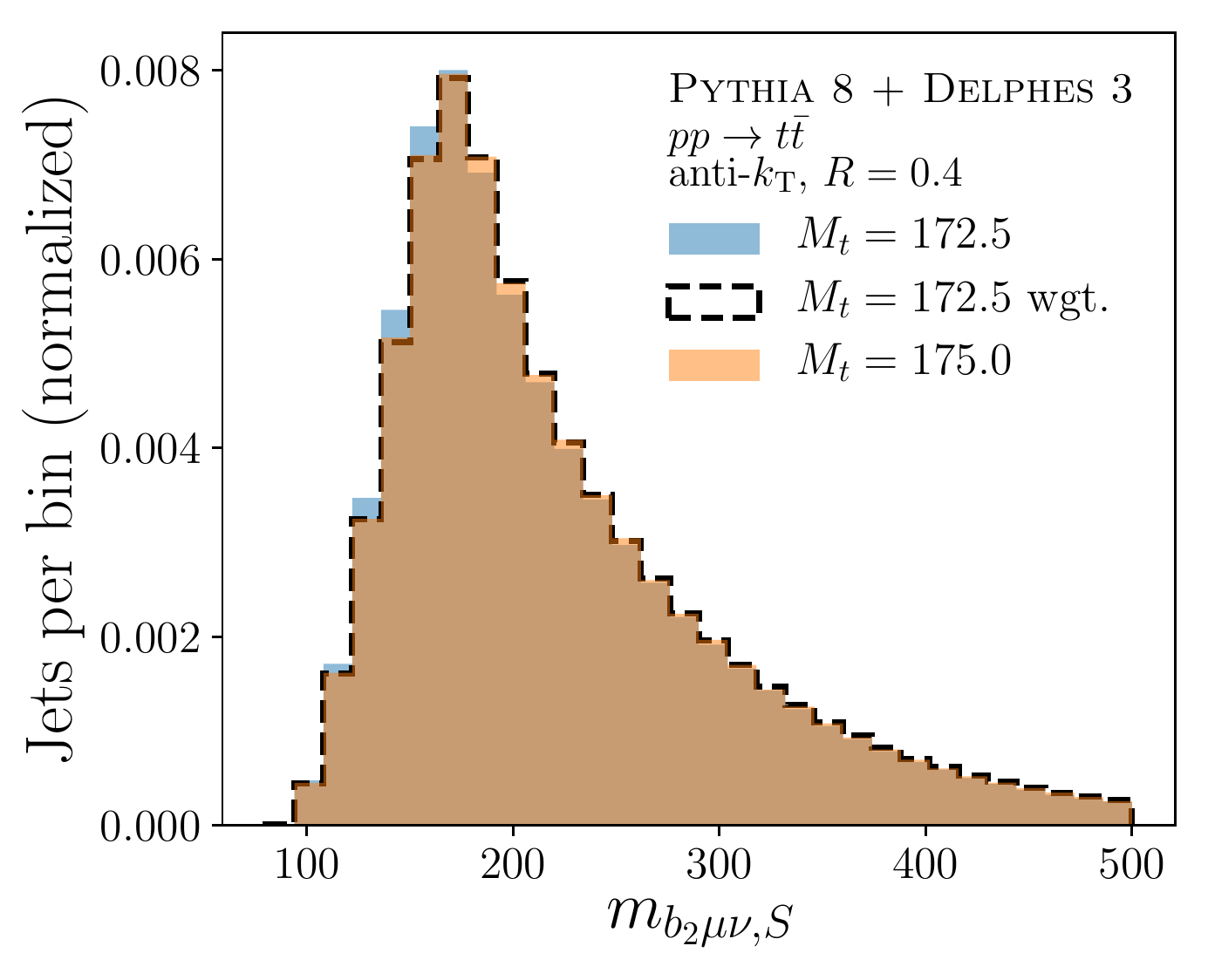}\\
	  
	  \includegraphics[scale=0.4]{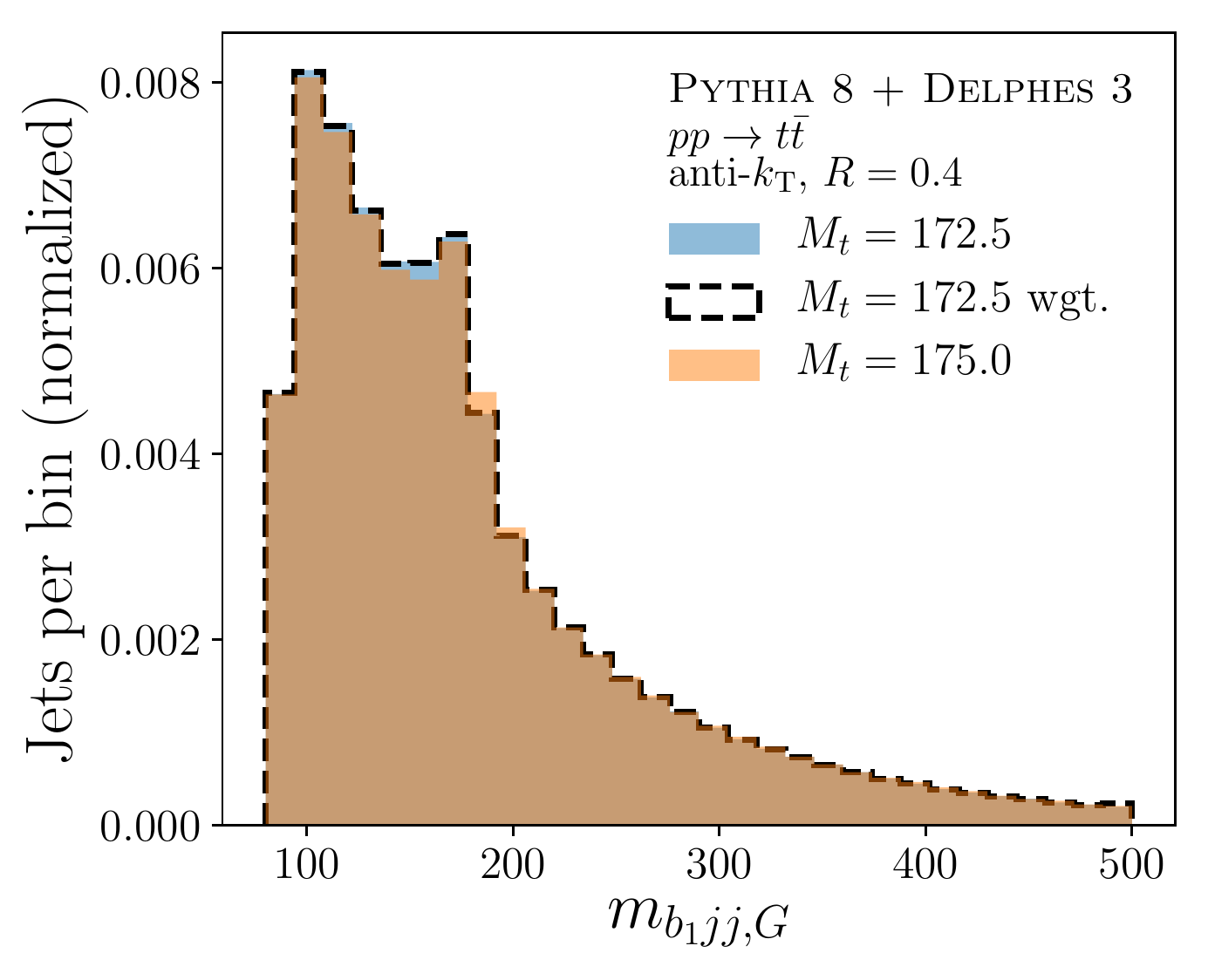}
	  \includegraphics[scale=0.4]{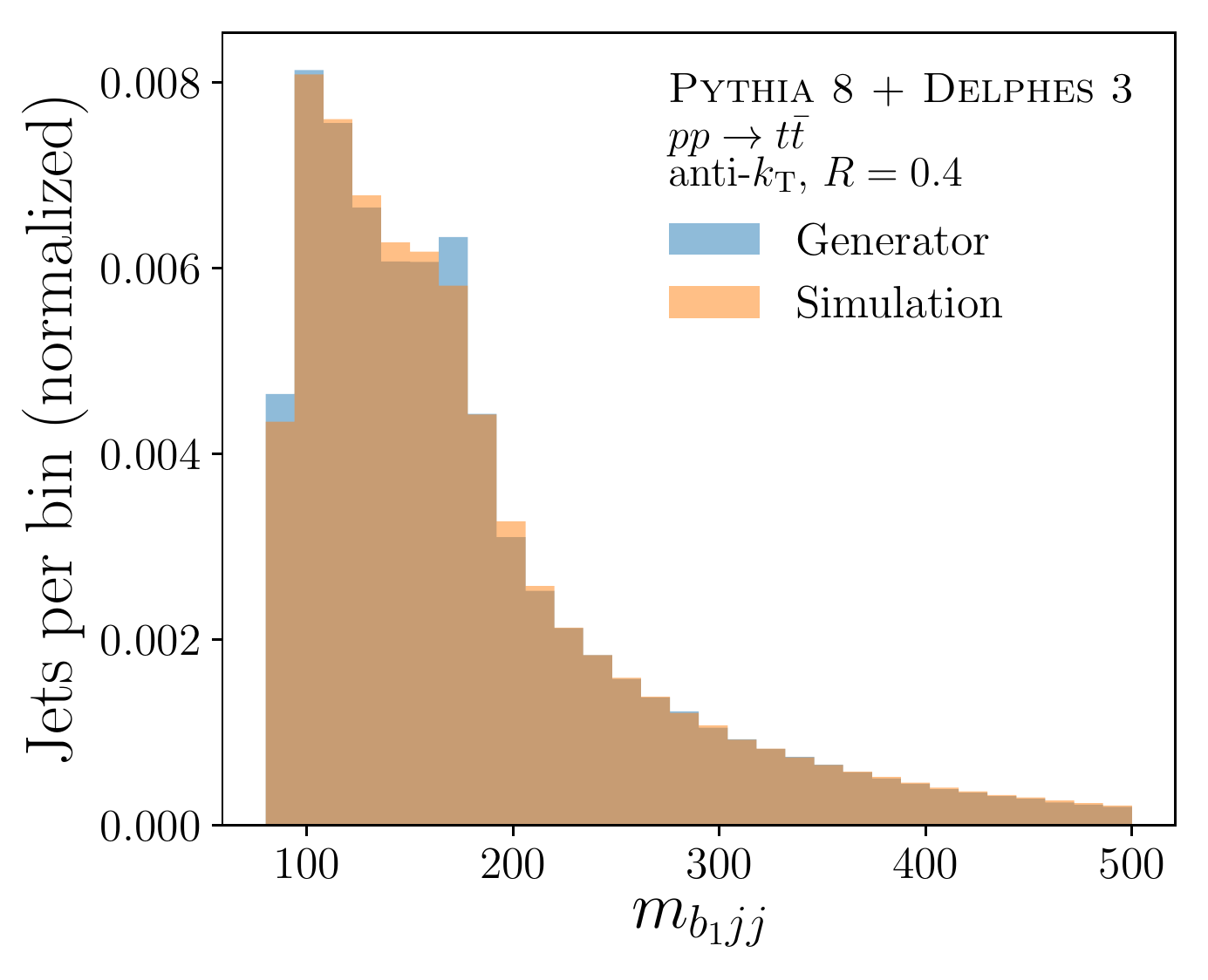}
	  \includegraphics[scale=0.4]{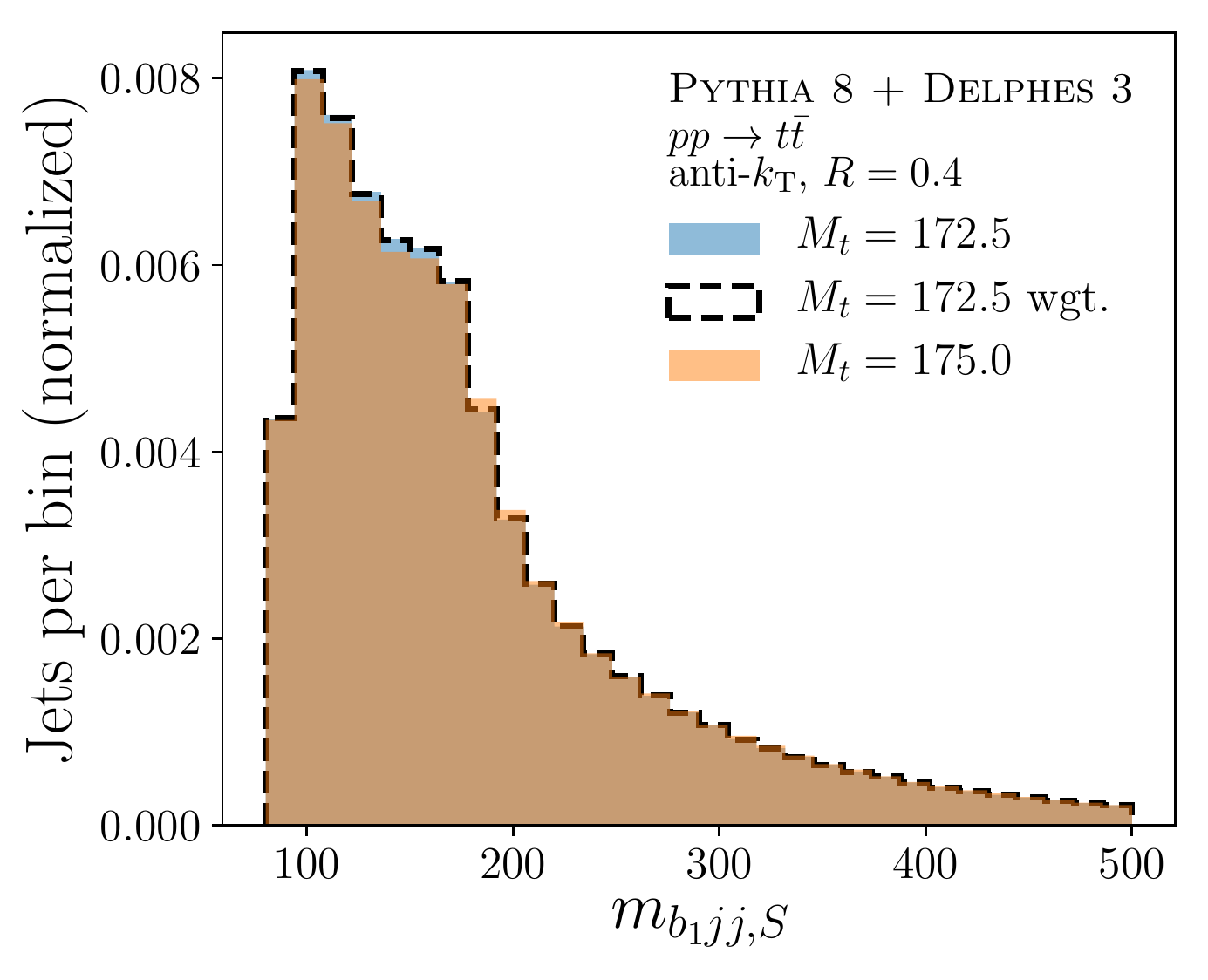}\\
    \caption{Histograms for the four observables for (left column) generator level for two different top quark masses, (middle column) a comparison of generator level and simulation level for a fixed top quark mass ($M_t = 172.5$ GeV), and (right column) simulation level for two different top quark masses.  Reweighted distributions are plotted over an average of 40 reweightings.}
    \label{fig:topfigures}
\end{figure*}

Histograms of the four observables for generator level and simulation level are presented in Fig.~\ref{fig:topfigures}.   On both particle and detector levels, one can see that varying the top quark mass $M_t$ has the greatest effect on $m_{b_1\mu\nu}$ and $m_{b_2\mu\nu}$ as opposed to $m_{b_2j_1j_2}$ and $m_{b_1j_1j_2}$.  However, the latter two still have some visible dependence on $M_t$.  Therefore, it is expected that fitting on all four observables (denoted $\mathcal{O}_4 = \{m_{b_1\mu\nu}, m_{b_2\mu\nu}, m_{b_2j_1j_2}, m_{b_1j_1j_2}\}$) should yield a more precise fit than fitting on any single one.

\begin{figure}[h!]
 \includegraphics[scale=0.41]{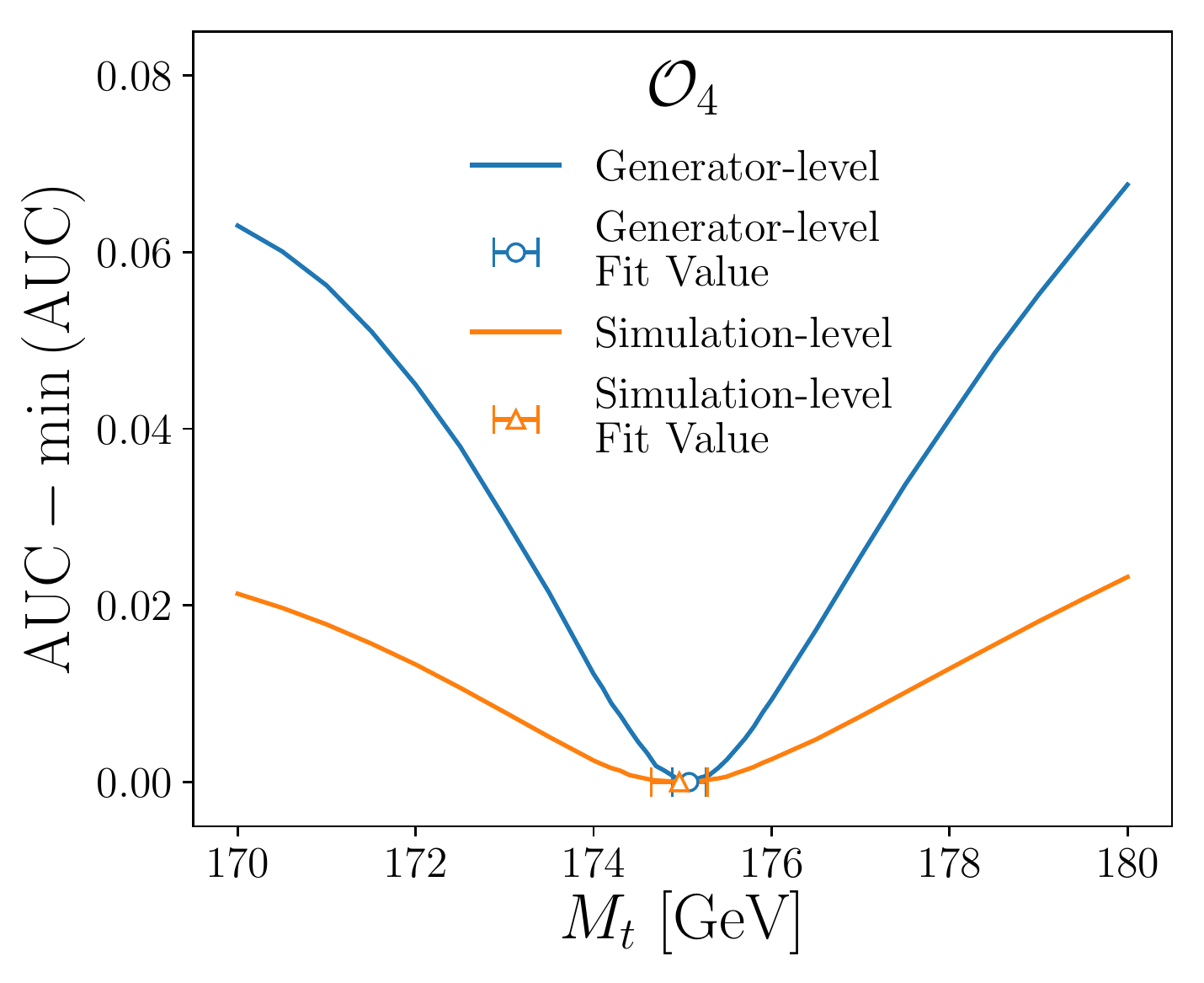}\\
 \includegraphics[scale=0.41]{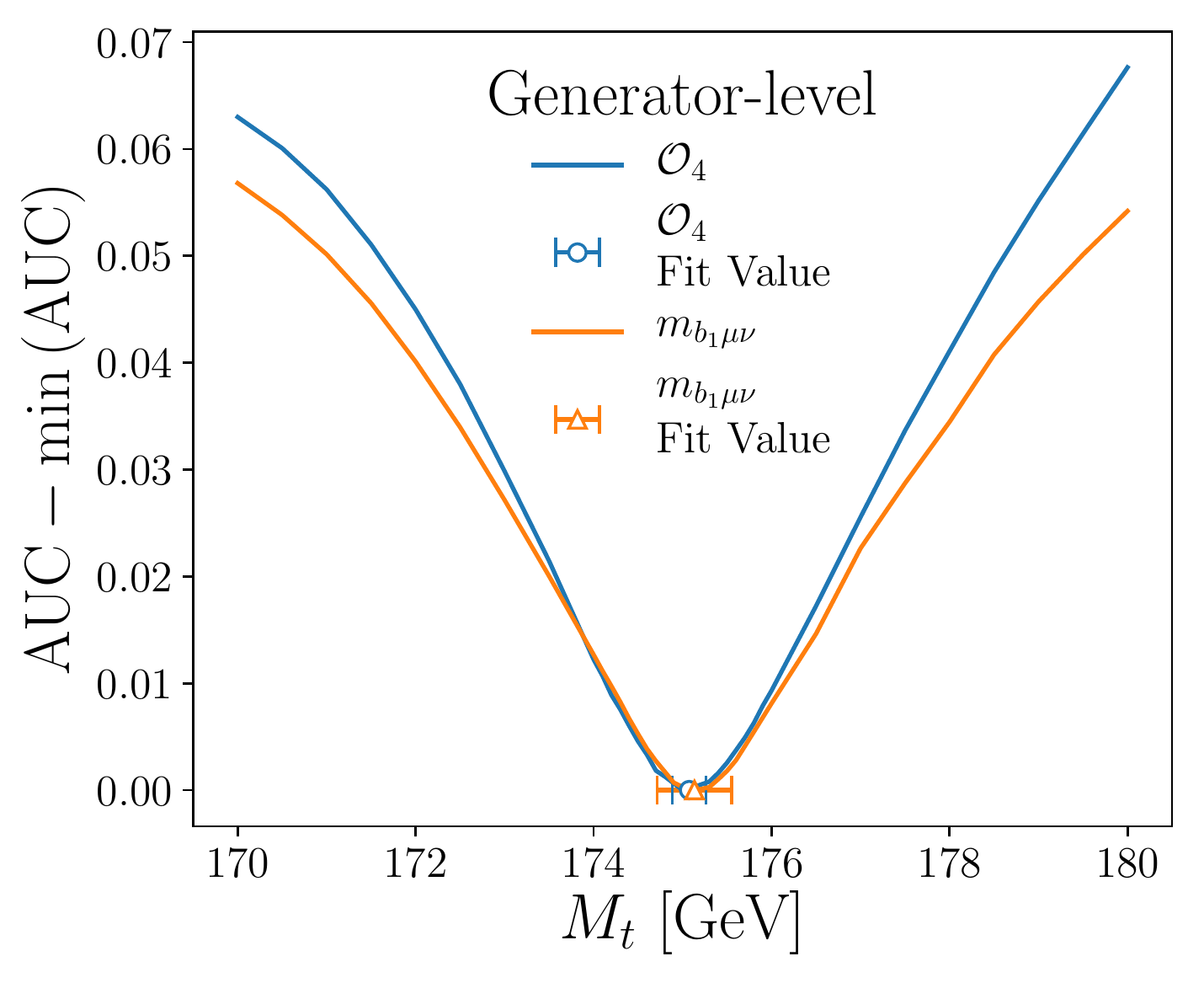}\\
 \includegraphics[scale=0.41]{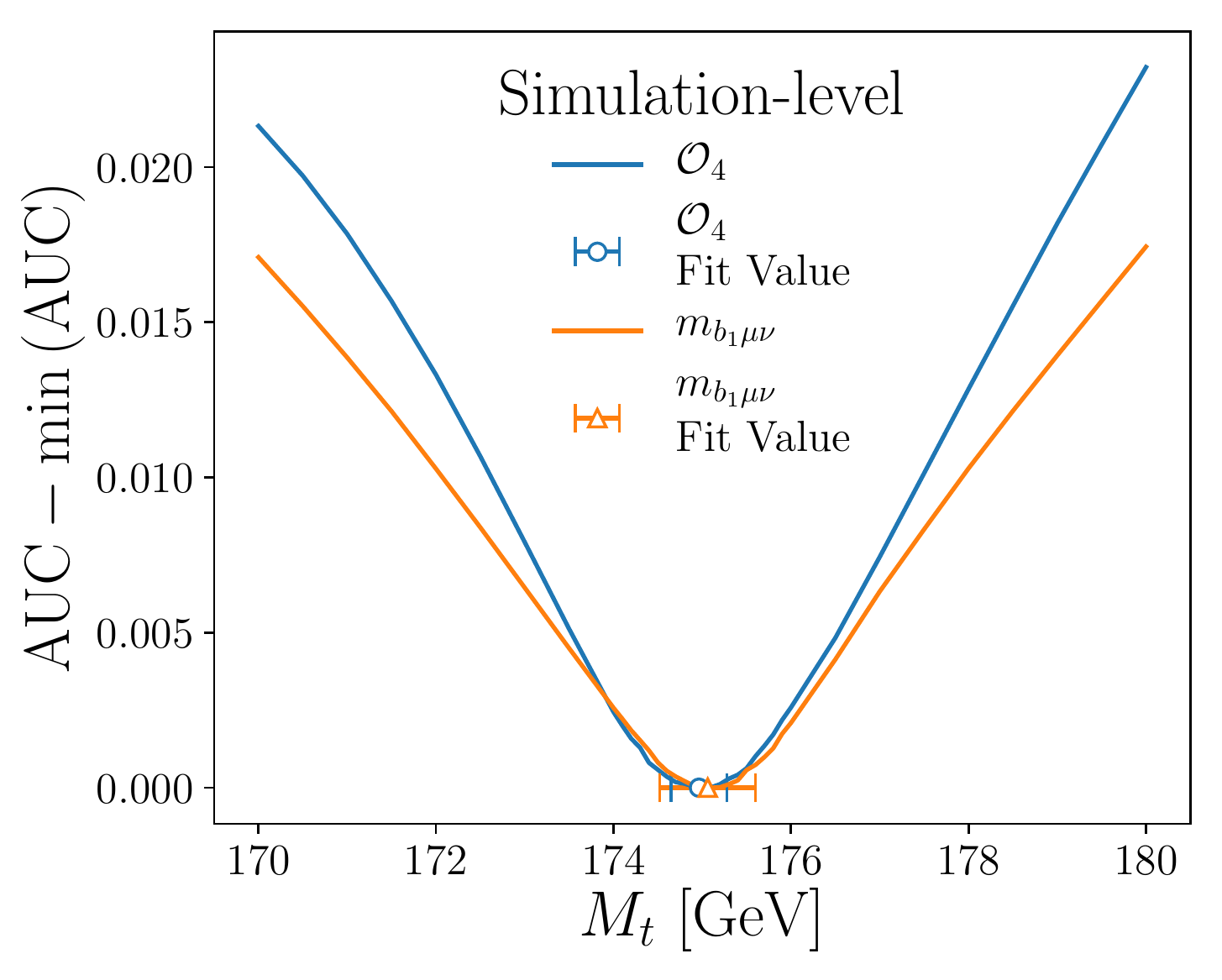}
    \caption{One-dimensional fits to the top quark mass.  The vertical axes show the increase of the AUC for the classifier $g$ from its minimum value.  The top plot uses all four observables and compares the fit at generator level to the fit at simulation level.  The middle (bottom) is at generator level (simulation level) and compares the fit with all four observables to the fit with only $m_{b_1\mu\nu}$.}
    \label{fig:topquarkfit}
\end{figure}
    
The application of the \textsc{Srgn} technique to the top quark mass fit is presented in Fig.~\ref{fig:topquarkfit}.  Both neural networks used for reweighting and classifying are implemented identically to the Gaussian example, with the exception of increasing early stopping patience to 20;   the training time is again about 5 seconds per epoch on an NVIDIA Tesla V100 GPU.  The parametrized reweighting is trained with $5.5 \cdot 10^6$ events with $M_t$ values sampled uniformly at random in the range [170 GeV, 180 GeV] against $1.4 \cdot 10^6$ events sampled with $M_t=172.5$ GeV.  To perform a given fit, we scan for the AUC as a function of the top quark mass with a step size of 0.1 GeV to search for the minimum.  There are $1.4 \cdot 10^6$ events in the nominal synthetic sample and $1.5 \cdot 10^6$ events in the fitting data.  In all cases, the fitted value agrees with the correct mass, $M_t=175$ GeV.  The top plot in Fig.~\ref{fig:topquarkfit} shows that the generator-level fit is much more precise than the simulation-level fit, based on the curvature of the AUC near the minimum.  The other two plots in the figure demonstrate a superior precision for the four-dimensional fit compared with the one-dimensional fit.  The same ensembling and outlier removal procedure is applied here as in the previous section.  Horizontal error bars are the standard deviation across 40 runs (outliers removed) with different random initializations.

Numerical values for the top quark mass fit are presented in Table~\ref{tab:topmassfit}.

\begin{table}[h!]
	\begin{tabular}{|c|c|c|c|c|} 
     \hline
     Parameter & Target [GeV]& Input & Level & Fit value [GeV]\\ 
     \hline
     \hline
     \multirow{4}{*}{$M_t$} & \multirow{4}{*}{175.00} & \multirow{2}{*}{$\mathcal{O}_4$} & Generator & 175.07 $\pm$ 0.19 \\
     & & & Simulation &  174.96 $\pm$ 0.31 \\
     \cline{3-5}
     & & \multirow{2}{*}{$m_{bl\nu}$} & Generator & 175.13 $\pm$ 0.42\\
     & & & Simulation & 175.06 $\pm$ 0.54 \\
     \hline
    \end{tabular}
    \caption{Numerical results for the top quark mass fit.  The reported values and errors represent the mean and standard deviation over the 40 runs (with outliers removed).}
        \label{tab:topmassfit}
\end{table}

\section{Conclusions and Outlook}
\label{sec:conclusions}

This paper addresses a key challenge with simulation-based inference in the presence of detector effects.  In particular, detector simulations are computationally expensive, so it is desirable to construct a method that uses as little detector simulation as possible.  We have introduced the \textsc{Srgn} approach that only requires one synthetic event sample with a detector simulation, and all other synthetic event samples need only be known at the generator level.  A variety of similar methods have been proposed in Ref.~\cite{Andreassen:2019nnm,Brehmer:2018kdj,Brehmer:2018eca,Brehmer:2019xox,Brehmer:2018hga}, but they typically require many synthetic event samples with detector simulation.

The \textsc{Srgn} protocol is unbinned and can process multidimensional feature spaces and parameter spaces.  In its current form, there is a nondifferentiable step required to optimize the area under the receiver operating characteristic curve.  Future refinements of this method may result in a fully differentiable pipeline.

\section*{Code and Data}

The code for this paper can be found at \url{https://github.com/hep-lbdl/SRGN}.  The synthetic data used for the numerical examples can be found at \url{https://zenodo.org/record/3518708} for the parton shower fits and at \url{https://zenodo.org/record/4067673} for the top quark mass fits.

\begin{acknowledgments}

This work was supported by the U.S. Department of Energy, Office of Science under Contract No. DE-AC02-05CH11231.  In particular, this work made use of the Cori supercomputer at the National Energy Research Scientific Computing Center (NERSC).  We are grateful to Wahid Bhimji for help with Cori-GPU.  This work was also supported by the NERSC Exascale Science Applications Program and the High Energy Physics Center for Computational Excellence.  S.-C. H. is supported by the U.S. Department of Energy, Office of Science, Office of Early Career Research Program under Award No. DE-SC0015971. We would like to thank Hannah R. Joo for suggesting the name of the algorithm and Anjali Chary for input on the schematic diagram. We also thank Gregor Kasieczka, Roman Kogler, Bryan Ostdiek, Reinhard Schwienhorst, and Jesse Thaler for feedback on the manuscript.

\end{acknowledgments}

\clearpage

\bibliography{myrefs}

\clearpage
\appendix

    \section{Weighted Loss Optimization}
    \label{app:derivation}
    
    This section derives Eq.~\ref{eq:g}, which is the optimal classifier function $g$, if using the weighted binary cross entropy loss function.  Given features $(X_G,X_S)$, labels $Y\in\{0,1\}$, weighting function $w$, and function $g$, a common loss functional to determine $g$ is the binary cross entropy:
    
    \begin{align}\nonumber
        \mathcal{L}[g]&=-Y\log(g(X_S))\\
        &\hspace{5mm}-(1-Y)w_{\theta_0}(X_G,\theta')\log(1-g(X_S))\,.
    \end{align}
    
    Conditioned on $X_S=x_S$, the expected loss is given by
    
    \begin{align}
    \nonumber
        \mathbb{E}[\mathcal{L}|X_S=x_s]&=-\mathbb{E}[Y|X_S=x_S]\log(g(x_S))\\\nonumber
        &\quad-\mathbb{E}[(1-Y)w_{\theta_0}(x_G,\theta')|X_S=x_S]\\\label{eq:conditionalexpectation}
        &\quad\quad \times\log(1-g(x_S))\,.
    \end{align}
    In general, $Y$ and $w(X_G)$ are not independent given $X_S$. However, we are assuming that $p(X_S|X_G)$ is the same in data and in simulation, so these two quantities should be approximately independent so long as the $X_G$ probability density is similar in data and in simulation.  With this approximation,
    
    \begin{align}
    \nonumber
        \mathbb{E}[\mathcal{L}|X_S=x_S]&\approx-\mathbb{E}[Y|X_S=x_S]\log(g(x_S))\\\nonumber
        &\quad-\mathbb{E}[1-Y|X_S=x_s]\\\label{eq:conditionalexpectation}
        &\quad\quad\times\mathbb{E}[w_{\theta_0}(x_G,\theta')|X_S=x_S]\log(1-g(x_S))\,.
    \end{align}
    
    By taking the derivative of Eq.~\ref{eq:conditionalexpectation} with respect to $g(x)$ and setting it equal to zero, one finds that
    
    \begin{align}
        g^*(x_S)\approx\frac{p}{\mathbb{E}[w_{\theta_0}(X_G,\theta')|X_S=x_S](1-p)+p}\,,
    \end{align}
    where since $Y$ is binary, $p\equiv \mathbb{E}[Y|X_S=x_S]=\Pr(Y=1|X_S=x_S)$.
    
    Furthermore, note that if $X_S=X_G$ (no detector effects), it is still the case that using $g^*(x)$ will generally not result in $\theta^*=\theta_?$.  However, if there is a loss function for $g$ such that when $g^*$ is inserted into the total loss, the result is Eq.~\ref{eq:dctrfit}, then $\theta^*=\theta_?$ (proven in Ref.~\cite{Andreassen:2019nnm}).  We do not know if such a function exists in general, but when $X_S\neq X_G$, this is irrelevant because $g^*$ cannot depend on $X_G$ as we do not have access to this information for the data.
    \section{Loss vs. AUC}
    \label{sec:app}

\begin{figure}[h!]
        \includegraphics[scale=0.5]{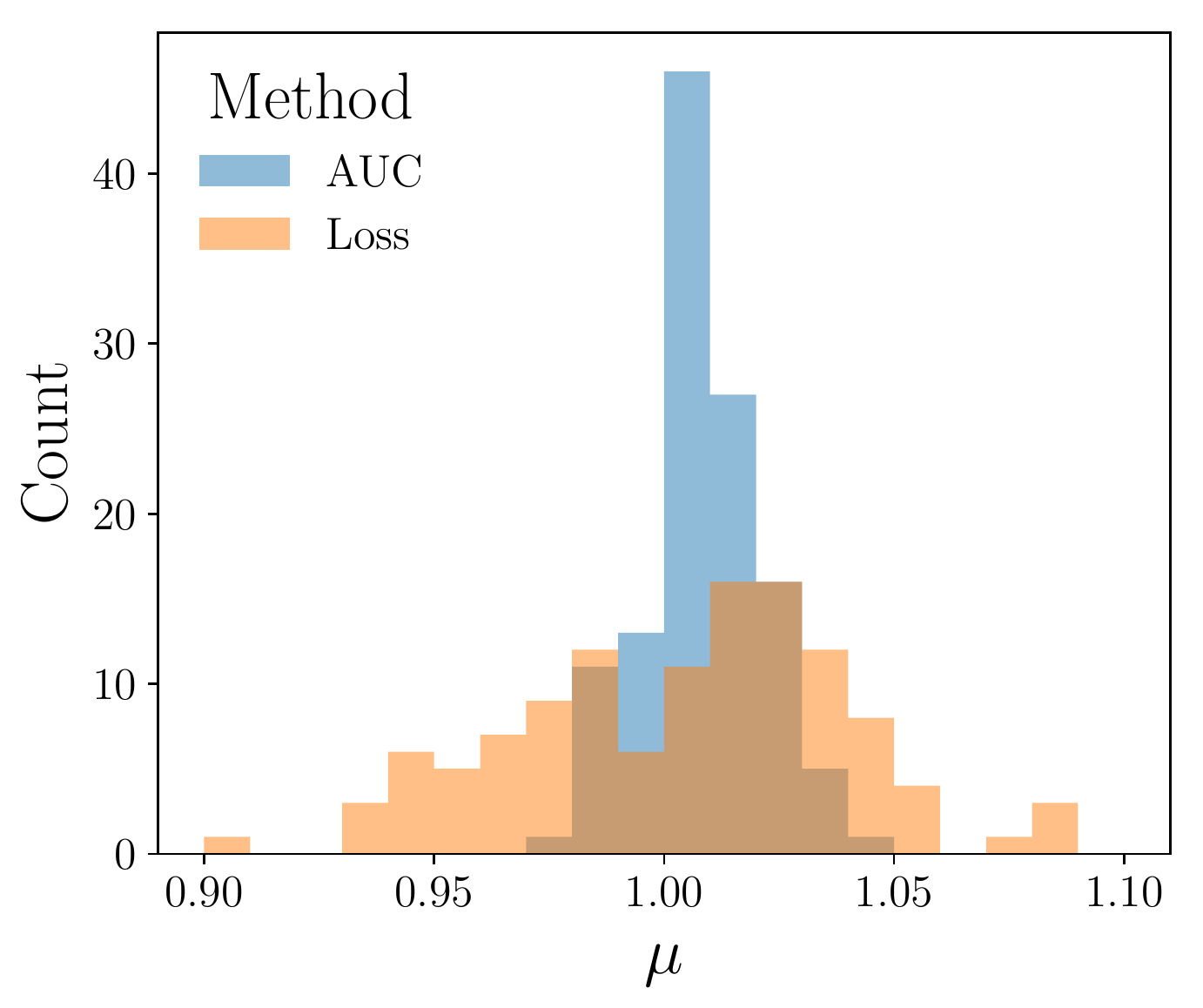}
    	   \\
        \caption{The distribution of fitted values $\mu_{\textsc{Srgn}}$ for the simple one-dimensional Gaussian case over 120 runs, comparing the methods of maximizing loss and minimizing AUC.}
        \label{fig:GaussianAUCvsLoss}
    \end{figure}

\begin{figure}[h!]
        \includegraphics[scale=0.5]{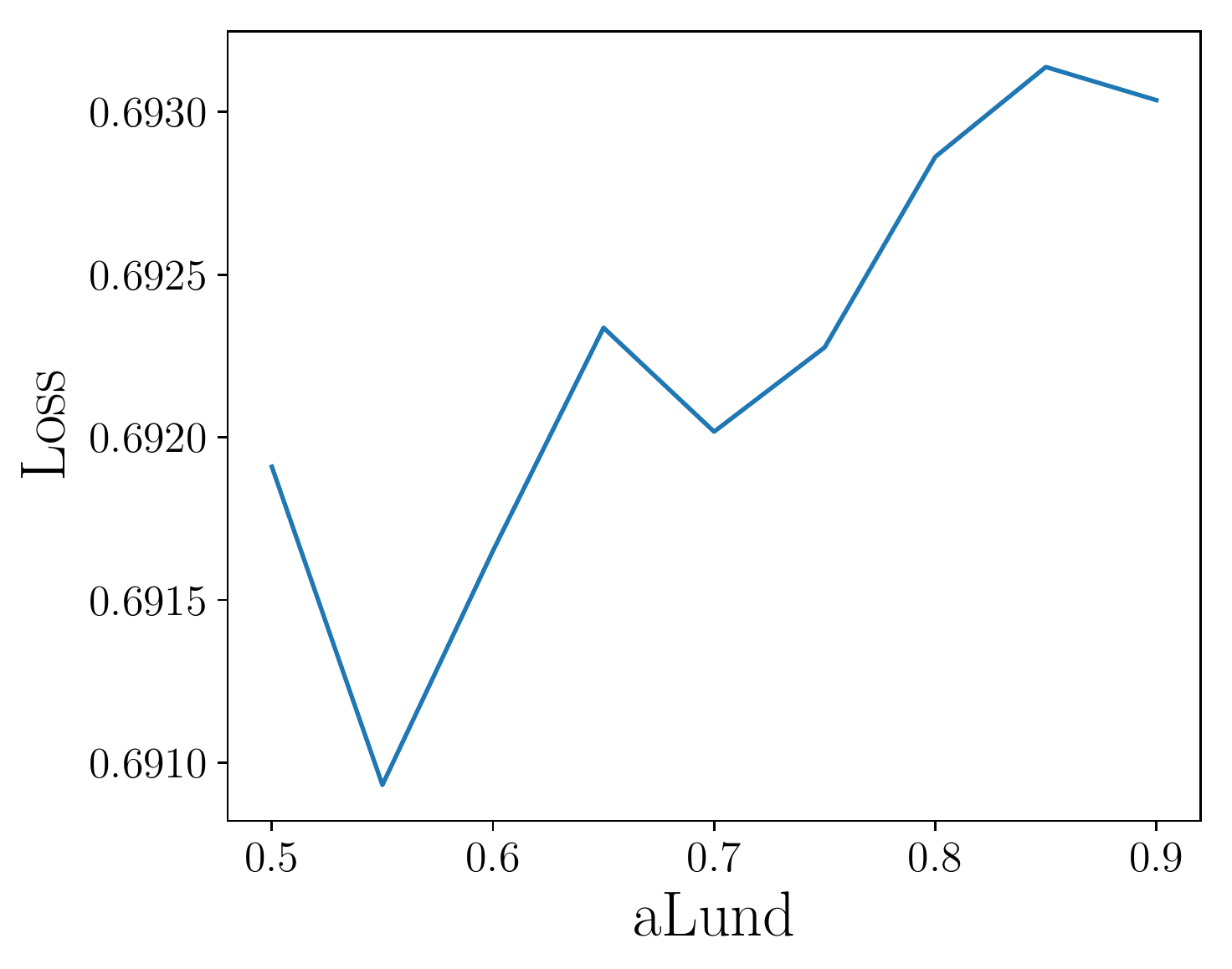}
    	   \\
        \caption{Plotted here are the fully trained loss values of the classifying step for various values of \lund, ensembled over 40 runs.  It is clear that the loss is not maximized for the correct value of \lund, 0.8000; conversely, AUC is (in comparison) smoothly minimized at the correct value (Fig.~\ref{fig:1dfitspartonshower}).}
        \label{fig:aLundLoss}
    \end{figure}

    As noted earlier, one may want to define 
    \begin{align}\nonumber
        &\theta_{\textsc{Srgn}}^* =\argmin_{\theta'}\max_{g}\left(\sum_{i\in\boldsymbol{\theta}_0}\log g(x_{S,i})\right.\\
        &\hspace{15mm}\left.+\sum_{i\in\boldsymbol{\theta_?}}\frac{f(x_{G,i},\theta')}{(1-f(x_{G,i},\theta'))}\log (1-g(x_{G,i}))\right)\,.
    \end{align}
    However, this generally does not reduce to $\theta^*=\theta_?$.  AUC still appears to be a more precise metric for parameter estimation even in the case where loss is employable, as illustrated in Fig.~\ref{fig:GaussianAUCvsLoss}. Furthermore, AUC is robust, whereas loss is unpredictable and unreliable for other parameters, as seen in Fig.~\ref{fig:aLundLoss}.
    
\end{document}